\documentclass[reprint,prd,nofootinbib,superscriptaddress]{revtex4}
\usepackage{CJK}
\usepackage{amsfonts}
\usepackage{amsmath}
\usepackage{amssymb}
\usepackage[english]{babel}
\usepackage{graphicx}
\usepackage{epsfig}
\usepackage{bm}
\usepackage{verbatim}
\usepackage[utf8]{inputenc}
\usepackage{booktabs}
\usepackage{multirow}
\usepackage{subfig}
\usepackage{slashed}
\usepackage{xcolor}
\usepackage[colorlinks=true,urlcolor=red,citecolor=red]{hyperref}
\usepackage[font=small]{caption}
\usepackage{float}
\usepackage{blindtext}
\usepackage{placeins}

\newenvironment{Eqnarray}%
{\arraycolsep 0.14em\begin{eqnarray}}{\end{eqnarray}}
\newcommand{\ba}{\begin{Eqnarray}}
	\newcommand{\ea}{\end{Eqnarray}}
\newcommand{\be}{\begin{equation}}
\newcommand{\ee}{\end{equation}}
\newcommand{\bal}{\begin{aligned}}
	\newcommand{\eal}{\end{aligned}}
\newcommand{\hs}{\hspace*{0.5cm}}
\renewcommand{\(}{\left(}
\renewcommand{\)}{\right)}
\renewcommand{\[}{\left[}
\renewcommand{\]}{\right]}
\newcommand{\abs}[1]{\left| #1 \right| }
\newcommand{\vev}[0]{VEV}
\newcommand{\vevs}[0]{VEVs}
\newcommand{\U}[1]{\mathrm{U}(1)_{\mathrm{#1}}}			
\newcommand{\SU}[2]{\mathrm{SU}(#1)_{\mathrm{#2}}}		

\newcommand{\mathsym}[1]{{}}

\definecolor{bostonuniversityred}{rgb}{0.8, 0.0, 0.0}

\input{tcilatex}
\begin{document}

\title{How low-scale Trinification sheds light in the flavour hierarchies, 
neutrino puzzle, dark matter and leptogenesis}

\author{A. E. C\'arcamo Hern\'andez}
\email{antonio.carcamo@usm.cl}
\affiliation{Universidad T\'ecnica Federico Santa Mar\'{\i}a 
and Centro Cient\'{\i}fico-Tecnol\'ogico de Valpara\'{\i}so\\
Casilla 110-V, Valpara\'{\i}so, Chile}

\author{D. T. Huong}
\email{dthuong@iop.vast.ac.vn}
\affiliation{Institute of Physics, VAST, 10 Dao Tan, Ba Dinh, Hanoi, Vietnam}

\author{Sergey Kovalenko}
\email{sergey.kovalenko@usm.cl}
\affiliation{Departamento de Ciencias F\'isicas, Universidad Andres Bello, \\
Sazi\'e 2212, Piso 7, Santiago, Chile}

\author{Ant\'onio P.~Morais}
\email{aapmorais@ua.pt}
\affiliation{Departamento de F\'{i}sica da Universidade de Aveiro and CIDMA
Campus de Santiago, 3810-183 Aveiro, Portugal}
\affiliation{Department of Astronomy and Theoretical Physics, \\
Lund University, Solvegatan 14A, SE-223 62 Lund, Sweden}

\author{Roman Pasechnik}
\email{Roman.Pasechnik@thep.lu.se}
\affiliation{Department of Astronomy and Theoretical Physics, \\
Lund University, Solvegatan 14A, SE-223 62 Lund, Sweden}

\author{Ivan Schmidt}
\email{ivan.schmidt@usm.cl}
\affiliation{Universidad T\'ecnica Federico Santa Mar\'{\i}a 
and Centro Cient\'{\i}fico-Tecnol\'ogico de Valpara\'{\i}so\\
Casilla 110-V, Valpara\'{\i}so, Chile}

\begin{abstract}
We propose a low-scale renormalizable trinification theory that successfully explains the flavor hierarchies and neutrino puzzle in the Standard Model (SM), as well as provides a dark matter candidate and also contains the necessary means for efficient leptogenesis. The proposed theory is 
based on the trinification $\SU{3}{C}\times \SU{3}{L}\times \SU{3}{R}$ gauge symmetry, which is supplemented 
with an additional flavor symmetry $\U{X}\times \mathbb{Z}_{2}^{(1)} \times \mathbb{Z}_{2}^{(2)}$. 
In the proposed model the top quark and the exotic fermions acquire tree-level masses, 
whereas the lighter SM charged fermions gain masses radiatively at one-loop level. In addition, 
the light active neutrino masses arise from a combination of radiative and type-I seesaw 
mechanisms, with the Dirac neutrino mass matrix generated at one-loop level. 

\footnotesize
DOI:\href{https://doi.org/10.1103/PhysRevD.102.095003}{10.1103/PhysRevD.102.095003}
\normalsize

\end{abstract}

\maketitle

\section{Introduction}

Despite the huge success of Standard Model (SM) as a theory of fundamental interactions, it has several open issues, 
which include the lack of explanation of the SM flavor structure, in particular, the fermion masses and mixing, 
the origin of Dark Matter (DM), as well as the source of parity violation in electroweak (EW) interactions. Besides, 
the SM features drawbacks such as the absence of sufficient CP violation and a strong departure from thermal 
equilibrium, both necessary for explaining the cosmological baryon asymmetry of the Universe. In this paper, 
we would like to address all these issues on the same footing in a single new framework, based on the trinification 
symmetry $\SU{3}{C}\times \SU{3}{L}\times \SU{3}{R}$ (see e.g. Refs.~\cite{Babu:1985gi,Choi:2003ag,Willenbrock:2003ca,Frampton:2004vw,Kim:2004pe,Carone:2004rp,Sayre:2006ma,Leontaris:2008mm,Cauet:2010ng,Stech:2012zr,Stech:2014tla,Hetzel:2015bla,Hetzel:2015cca,Pelaggi:2015kna,Dong:2017zxo,Wang:2018yer,Dinh:2019jdg}) supplemented with an additional flavor symmetry, $\U{X}\times \mathbb{Z}_{2}^{(1)} \times \mathbb{Z}_{2}^{(2)}$, 
with the spontaneously broken $\U{X}\times \mathbb{Z}_{2}^{(1)}$ symmetry (we refer to this model as to 3331 framework 
in what follows). We have included a preserved $\mathbb{Z}_{2}^{(2)}$ symmetry to implement the one-loop level 
radiative seesaw mechanisms that generate the light active neutrino masses as well as the masses for the SM charged 
fermions below the top quark mass. The spontaneously broken $\mathbb{Z}_{2}^{(1)}$ symmetry allows us to separate 
the scalar bi-triplet that gives tree-level masses for the top and exotic up-type quarks from the one that produces 
the exotic down-type quark masses.

Our 3331 model is the most economical theory of trinification that can naturally explain the masses and hierarchies 
for the SM fermions, since unlike the other existing trinification-based models such as the one presented in 
Ref.~\cite{Dong:2017zxo}, it does not rely on the inclusion of a large variety of additional representations such 
as scalar anti-sextets in order to generate the light active neutrino masses in the SM. In our case, 
the actual mechanism for the light neutrino mass generation is provided by a one-loop level radiative seesaw. 
Moreover, our model is capable of simultaneously explaining the hierarchy of charged SM fermion masses, 
which is not considered in earlier works. We also investigate the potential implications of this scenario for DM 
and leptogenesis.

\section{The model}
\label{Sec:Model}

Our model is based on the $\SU{3}{C}\times \SU{3}{L}\times \SU{3}{R}\times \U{X}$ gauge symmetry which
can be motivated by a high-scale ultraviolet (UV) completion based on the embedding of trinification 
as a subgroup of ${\rm E}_6$ \cite{Gursey:1975ki,Shafi:1978gg,Stech:2003sb}, whereas an additional flavor 
symmetry can be inspired by the coset of the ${\rm E}_8 \to [\SU{3}{}]^3$ reduction
\cite{Camargo-Molina:2016bwm,Camargo-Molina:2016yqm,Camargo-Molina:2017kxd,Morais:2020odg,Morais:2020ypd}. 
Indeed, in the framework of a super-string inspired $\mathbb{Z}_{\rm n}$ orbifolding procedure of ${\rm E}_8$ reduction \cite{Katsuki:1989kd}, 
one of the possible ${\rm E}_8$ breaking patterns features the following scheme 
\begin{eqnarray}
{\rm E}_8 \to \SU{3}{C}\times \SU{3}{L}\times \SU{3}{R}\times [\U{}]^2 \,,
\end{eqnarray}
where the rank-reduction $[\U{}]^2 \to \U{X}$ can occur
via a Higgs mechanism. All the subsequent breaking steps may in principle take place at any energy scale between the ${\rm E}_8$, 
or alternatively ${\rm E}_6$, ($M_\mathrm{GUT}$) and the EW ($M_{\rm EW}$) breaking scales. Here, we consider a particularly 
attractive opportunity of the trinification breaking at a relatively low scale compared to the ${\rm E}_8$ breaking one, 
$M_\mathrm{GUT} \ggg M_3 \sim 100$ TeV, i.e. not too far from the reach of future collider measurements.

In fact, both supersymmetric and non-supersymmetric realizations of this model involving the $\rm{ E}_6$ symmetry would 
require a very high scale for both the ${\rm E}_6$ and the trinification breaking scales, above $10^{16}$ GeV, due to strong constraints 
on the ${\rm E}_6$ gauge mediated proton decay. In order to relax this constraint, we explore a non-supersymmetric version 
of the model, without manifest embedding of its particle content into representations of ${\rm E}_6$ or even ${\rm E}_8$, 
such that the scale of the trinification symmetry breaking can indeed be within the reach of future colliders such as 
the $100$ TeV proton-proton Future Circular Collider (FCC).

Our model realizes the following particular symmetry breaking scheme:
\begin{eqnarray}
&&\mathcal{G} \equiv \SU{3}{C}\times \SU{3}{L}\times \SU{3}{R}\times \U{X}\times 
\mathbb{Z}_{2}^{(1)}\times \mathbb{Z}_{2}^{(2)}
{\xrightarrow{v_{\chi}^{(k)}, v_{\chi}^{(4)}}}  \notag \\
&&\hspace{7mm}
\SU{3}{C}\times \SU{2}{L}\times \SU{2}{R}\times \U{B-L}\times \mathbb{Z}_{2}^{(2)}
{\xrightarrow{w_{\chi}}}  \notag \\
&&\hspace{7mm}
\SU{3}{C}\times \SU{2}{L}\times \U{Y}\times \mathbb{Z}_{2}^{(2)}
{\xrightarrow{v}}  \notag \\
&&\hspace{7mm}
\SU{3}{C}\times \U{Q}\times \mathbb{Z}_{2}^{(2)} \,,  
\label{SB}
\end{eqnarray}
where the different symmetry breaking scales fulfill the following hierarchies:
\begin{equation}
v \ll w_{\chi } \ll v_{\chi}^{(k)} \sim v_{\chi}^{(4)} \sim M_3 \,, 
\hspace{1.5cm} k=1,2 \,.
\label{VEVsinglets}   
\end{equation}
Here, $v=246$ GeV is the EW symmetry breaking scale. We assume that the trinification 
breaking scales $v_{\chi}^{(k)}$ and $v_{\chi}^{(4)}$ are of the order of $100$ TeV, 
which would make our model potentially testable at the future FCC $100$ TeV collider.

In our model the exotic particles carry the SM electric charge which is defined in terms of the trinification symmetry 
generators as follows:
\begin{equation}
Q = T_{\rm 3L} + T_{\rm 3R} + \beta (T_{\rm 8L} + T_{\rm 8R}) + X = 
T_{\rm 3L} + T_{\rm 3R} + \frac{1}{2}(B-L) \,,
\label{eq:def-Q}
\end{equation}
where $\beta$ and the baryon minus lepton number operator $B-L$ are defined as
\begin{equation}
\beta = -\frac{1}{\sqrt{3}} \,, \qquad B-L = 2[ \beta ( T_{\rm 8L} + T_{\rm 8R} ) + X ] \,.
\label{eq:def-beta} 
\end{equation}
We have chosen such particular value of $\beta$ in order to have the third component 
of the leptonic triplet to be electrically neutral, which allows for a consistent 
implementation of a low-scale seesaw mechanism for light active neutrino masses generation.

The scalar sector of our model is composed only of three scalar bi-triplets and one 
$\SU{3}{L}$ scalar triplet that feature the following patterns of the vacuum 
expectation values (\vevs)
\begin{eqnarray}
\langle \chi_{1}\rangle &=&\left( 
\begin{array}{ccc}
\frac{v}{\sqrt{2}} & 0 & \frac{w_{\chi}}{\sqrt{2}} \\ 
0 & 0 & 0 \\ 
0 & 0 & \frac{v_{\chi }^{(1)}}{\sqrt{2}}
\end{array}
\right) \,,
\hspace{1cm}\hspace{1cm}
\langle \chi_{2}\rangle =\left( 
\begin{array}{ccc}
0 & 0 & 0 \\ 
0 & 0 & 0 \\ 
0 & 0 & \frac{v_{\chi }^{(2)}}{\sqrt{2}}
\end{array}
\right) \,, \notag \\
\langle \chi_{3}\rangle &=&\left( 
\begin{array}{ccc}
0 & 0 & 0 \\ 
0 & 0 & 0 \\ 
0 & 0 & 0%
\end{array}
\right) \,,
\hspace{1cm}\hspace{1cm}
\langle \chi_{4}\rangle =\left( 
\begin{array}{ccc}
0 & 0 & \frac{v_{\chi}^{(4)}}{\sqrt{2}}
\end{array}
\right) \,,
\label{VEVs-tri}
\end{eqnarray}
where $\chi_{3}$ does not acquire a \vev since it is charged under the
preserved $\mathbb{Z}_{2}^{(2)}$ symmetry. The latter has been incorporated in order to implement 
the one-loop level radiative seesaw mechanisms for producing the light active neutrino masses, 
as well as the masses for the SM charged fermions lighter than the top quark. 

A justification of the chosen structure of the scalar sector is as follows. The scalar bi-triplet
$\chi_1$ is needed to generate tree-level masses for the top and exotic up-type quarks, whereas the
scalar bi-triplet $\chi_2$ is required to produce tree-level masses for the exotic down-type quarks.
We include the spontaneously broken $\mathbb{Z}_{2}^{(1)}$ symmetry in order to separate the scalar
bi-triplets $\chi_1$ and $\chi_2$ in the mass spectrum. Furthermore, the inclusion of the scalar
bi-triplet $\chi_3$ is needed for the implementation of the one-loop level radiative seesaw 
mechanism that generates the masses for the SM charged fermions below the top quark mass scale. 
Note that the presence of potentially light pseudo-Goldstone
 CP-odd state in the scalar spectrum of our model offers a long-lived scalar candidate for warm DM, known 
in the literature as the Majoron -- an appealing feature of the considered model further discussed below in Sec.~\ref{Sec:DM}.
In addition, the scalar bi-triplet $\chi_3$ is crucial to generate the masses for the light 
active neutrinos. Besides, the scalar triplet $\chi_4$ is necessary to generate the quark mixing 
angles in the 13 and 23 planes.

It is worth mentioning that the lightest $125$ GeV SM-like Higgs boson is mainly composed of the 
CP-even neutral component of $(\chi_{1})_{11}$. Moreover, our model can naturally accommodate 
its alignment limit since all the other scalar states are typically very heavy and thus are decoupled 
from the SM in the mass spectrum, as detailed below in Sec.~\ref{Sec:scalar-sector}. This also implies 
the absence of tree-level flavour changing neutral currents (FCNCs) for the SM-like Higgs boson state, 
while such contributions from the heavier scalars are strongly suppressed by their large mass scale.

The fermion sector of the model is motivated by the conventional 
331 model and is obtained by the Left-Right (LR) symmetrization, which leads to the following multiplet structure: 

\begin{equation}
Q_{n(L,R)}=
\begin{pmatrix}
d_{n} \\ 
-u_{n} \\ 
J_{n}%
\end{pmatrix}
_{L,R} \,,
\hspace{1cm}
Q_{3L}=
\begin{pmatrix}
u_{3} \\ 
d_{3} \\ 
U%
\end{pmatrix}
_{L,R} \,,
\hspace{1cm}
L_{\alpha(L,R)}=
\begin{pmatrix}
\nu_{\alpha} \\ 
e_{\alpha} \\ 
N_{\alpha}
\end{pmatrix}
_{L,R} \,,\hspace{1cm}n=1,2, \hspace{1cm} \alpha =1,2,3 \,.
\end{equation}
The model predicts exotic leptons $N_{\alpha L,R}$ and quarks $J_{\alpha L,R}$, apart from the right-handed neutrinos $\nu_{\alpha R}$.  
The transformation properties of the scalar and fermionic fields under the
symmetries of the model are shown in Tables~\ref{tab:scalars} and 
\ref{tab:fermions}. Note that with the above fermion multiplet structure the model is anomaly-free. The anomaly cancelation conditions are analogous to that in the 331 models 
\cite{Pisano:1991ee,Frampton:2004vw,Foot:1992rh,Singer:1980sw,Montero:1992jk,Foot:1994ym,Dong:2006mg,Boucenna:2015zwa,Valle:2016kyz}
and require that both for   
$\SU{3}{L}$ and $\SU{3}{R}$ the number of fermion generations match the number of colors and the third quark generation has different transformation properties under $\SU{3}{L,R}$ than the first two generations.

With the previously specified particle content and symmetries, the Lagrangian for Yukawa interactions reads as:
\begin{equation}
-\mathcal{L}_{Y}^{(Q) }=y_{3}^{(Q) }\overline{Q}_{3L}\chi _{1}Q_{3R} +
\sum_{n=1}^{2}\sum_{m=1}^{2}y_{nm}^{(Q)}
\overline{Q}_{nL}\chi_{2}^{\ast}Q_{mR} + \sum_{n=1}^{2} y_{n3}^{(Q)}
\overline{Q}_{nL}\chi_{3}^{\ast}Q_{3R} + {\rm h.c.} \,,
\label{Lq}
\end{equation}
\begin{equation}
-\mathcal{L}_{Y}^{(l)} = \sum_{\alpha =1}^{3}\sum_{\beta=1}^{3}
(x_{\chi})_{\alpha \beta }\overline{L}_{\alpha L}\chi_{2}L_{\beta R}+
\sum_{\alpha =1}^{3}\sum_{n=1}^{2}( x_{\Psi })_{\alpha n}\overline{L}_{\alpha L}
\chi_{4}\Psi_{nR}+\sum_{n=1}^{2}\sum_{m=1}^{2}(m_{\Psi})_{nm}\Psi_{nR}
\overline{\Psi_{mR}^{c}} + {\rm h.c.} \,.
\label{Ll}
\end{equation}

\begin{figure}
    \centering
    \includegraphics[width = 0.9\textwidth]{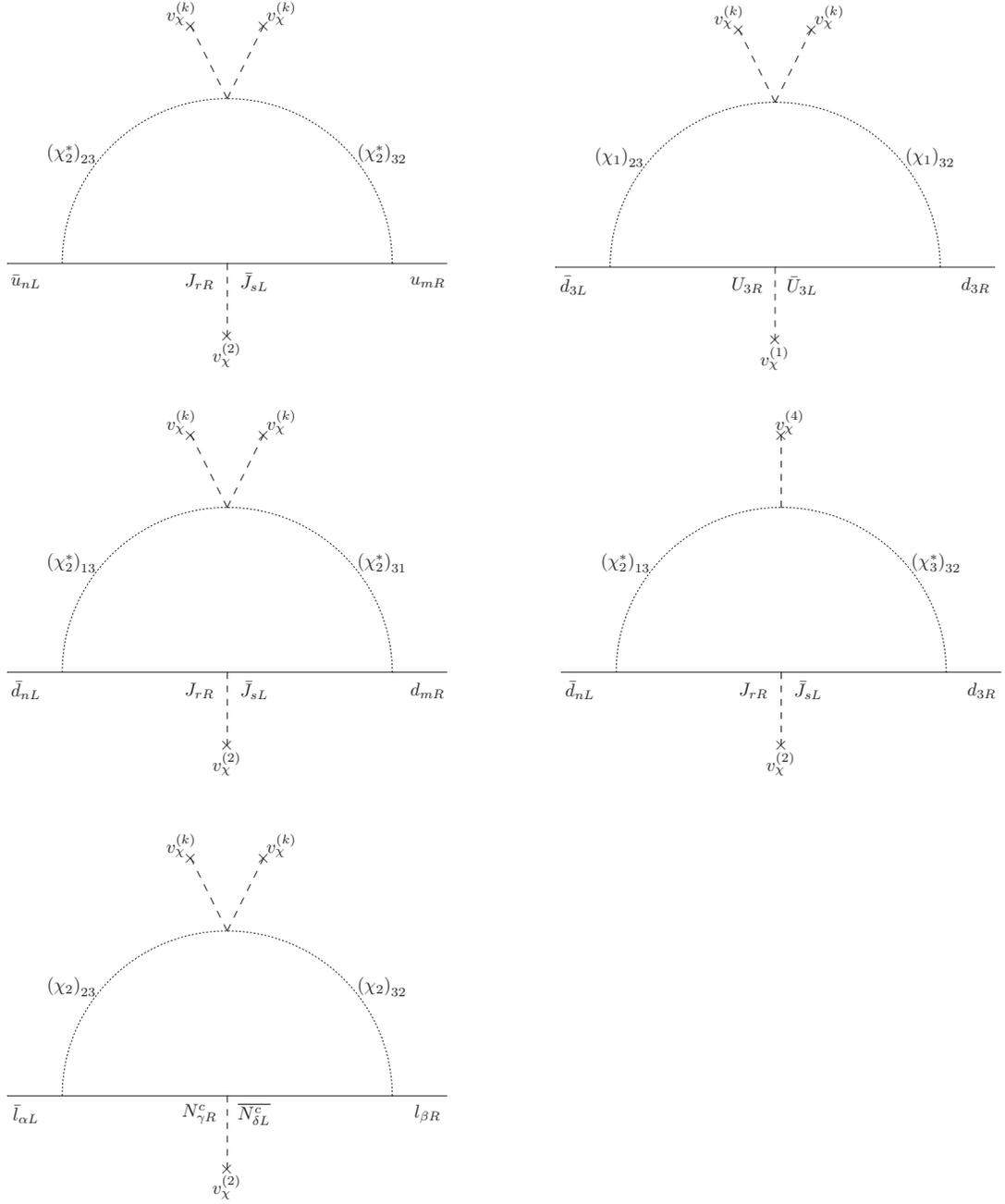}
    \caption{One-loop Feynman diagrams contributing to the entries of the SM
charged fermion mass matrices. Here, $n,m,k,r,s=1,2$ and $\protect\alpha,
\protect\beta,\protect\gamma,\protect\delta=1,2,3$.}
\label{Loopdiagramsq}
\end{figure}

\begin{figure}
    \centering
    \includegraphics[width = 0.9\textwidth]{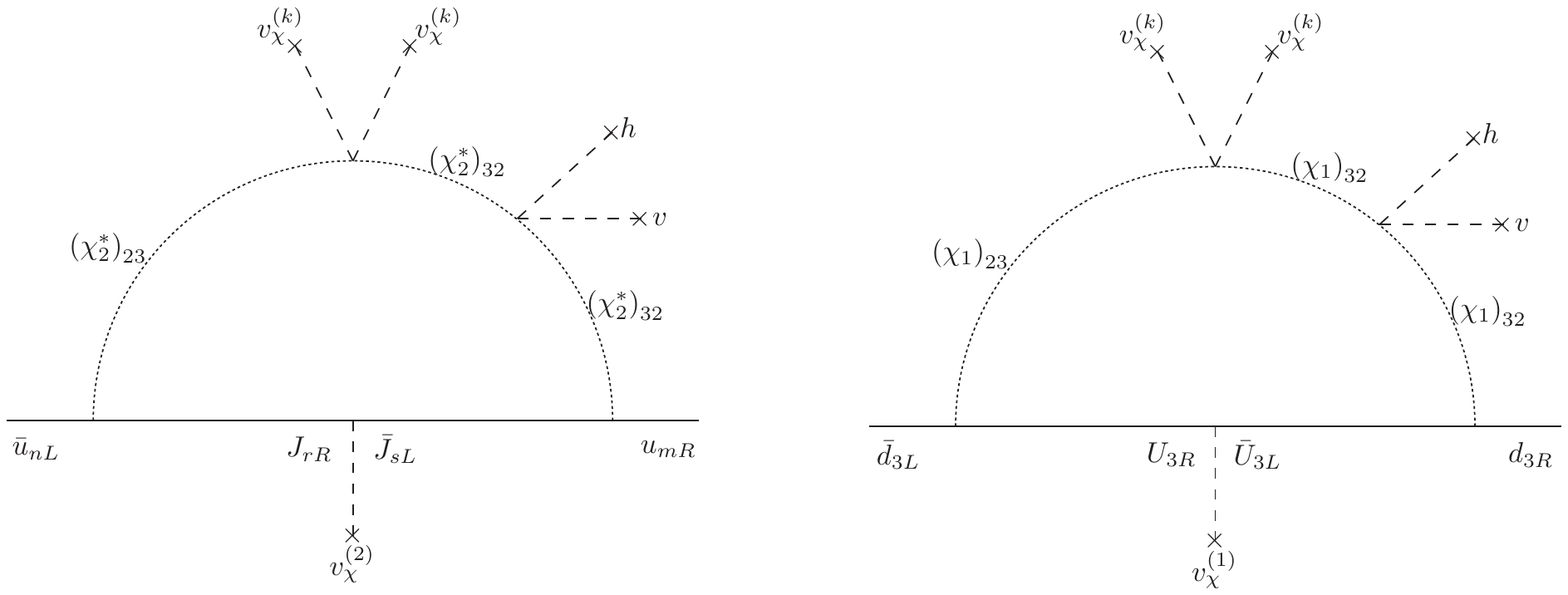}\vspace{-12cm}
    \caption{Some one-loop Feynman diagrams contributing to the $hf\bar{f}$ interactions. Here, $n,m=1,2$, $h$ is the $125$ GeV SM-like Higgs boson and $f$ a SM fermion lighter than the top quark.}
\label{Loopdiagramshffbar}
\end{figure}

If follows from \eqref{Lq} and \eqref{Ll} that the top and
exotic up- and down-type quarks get tree-level masses which are directly proportional to $v$, $v_{\chi}^{(1)}$ and $v_{\chi}^{(2)}$, respectively. 
The mixing between the top and the exotic $U$ quark is controlled by 
the $w_{\chi}$ \vev. In addition, after the first step of the trinification 
symmetry breaking, the exotic, vector-like, quarks remain singlets under 
the LR $\SU{2}{L}\times \SU{2}{R}$ symmetry group. Furthermore, the SM charged fermions, which are significantly lighter than the top quark, 
get one-loop masses via a radiative seesaw mechanism mediated by heavy exotic fermions 
and heavy scalars running in the internal lines of the loop, as shown in Fig.~\ref{Loopdiagramsq}. Notice, as shown in Fig.~\ref{Loopdiagramsq}, that the one-loop masses for the light SM charged fermions 
receive contributions from $v_{\chi}^{(1)}$ and $v_{\chi}^{(2)}$. However, it is worth mentioning 
that the loop-generated masses for the up, charm, down and strange quarks and the SM charged leptons 
are mainly fixed by $v_{\chi}^{(2)}$ \vev, whereas the bottom quark mass is mainly determined by 
$v_{\chi}^{\left(1\right) }$ \vev. This is due to the fact that the charged exotic fermions that 
induce such one-loop masses are proportional to those \vevs. In addition, the Cabbibo mixing 
is mainly fixed by $v_{\chi}^{(2)}$, since it is generated by the one-loop contribution 
mediated by the exotic down-type quarks. Notice that the Cabbibo mixing receives contributions 
from both up- and down-type quark sectors, whereas the down-type quark sector helps to generate
the remaining mixing angles. The \vevs~that are crucial for generating the quark mixing angles in 
the 13 and 23 planes are both $v_{\chi}^{(2)}$ and $v_{\chi}^{(4)}$, as indicated in Fig.~\ref{Loopdiagramsq}. As seen from Fig.~\ref{Loopdiagramsq}, where the one loop Feynman diagrams have Trinification VEVs in the external legs, it follows that the dominant contribution for the masses of such SM-like charged fermions is not due the Electroweak Symmetry Breaking (EWSB) mechanism but mostly due to the trinification breaking. Despite of this issue, the interactions of the $125$ GeV SM like Higgs boson $h$ with SM fermion-antifermion pairs, such as $hb\bar{b}$, $hc\bar{c}$ can be effectively generated at one loop level via the quartic scalar interaction insertions $\left(\chi_1\chi^{\dagger}_1\right)^2$ and $\left(\chi_1\chi^{\dagger}_1\right)\left(\chi_2\chi^{\dagger}_2\right)$ where the EW scale VEV and the $125$ GeV SM like Higgs boson $h$ are in the external legs, in addition to the Trinification VEVs. Some of the one-loop Feynman diagrams contributing to the $hf\bar{f}$ interactions are shown in Fig.~\ref{Loopdiagramshffbar}. Furthermore, despite of the fact that the dominant contribution for the masses of the SM-like charged fermions lighter than the top quark is mostly due to the trinification breaking, the masses of such SM charged fermions can be successfully reproduced by having appropiate values of the  masses of non SM scalars and exotic fermions running in the internal lines of the loops and of the quartic scalar couplings and Yukawa couplings. For instance, to succesfully explain the GeV scale value of the bottom quark and tau lepton masses, from Fig.~\ref{Loopdiagramsq}, we have that such masses can be estimated as:
\begin{equation}
m_b\sim m_{\tau}\sim\frac{y^2}{16\pi^2}\lambda\frac{v^2_{\chi}}{M_{\rm F}},
\label{estimate}
\end{equation}
where $M_{\rm F}$ is the mass scale of the exotic fermions, $y$ the SM fermion-exotic fermion Yukawa coupling and $\lambda$ the quartic scalar coupling. Taking $v_{\chi}\sim M_{\rm F}\sim\mathcal{O}\left(100\right)$ TeV and $y\sim \lambda \sim \mathcal{O}\left(0.1\right)$, Eq. (\ref{estimate}) takes the form $m_b\sim m_{\tau}\sim 10^{-5}v_{\chi}\sim\mathcal{O}\left(1\right)$ GeV, thus showing that our model naturally explains the smallness of the bottom and tau masses with respect to the EWSB scale. Furthermore, despite the fact that the masses of the light SM charged fermions (below the top quark mass) 
are generated at one-loop level, the hierarchy between such masses can be accommodated by having 
some deviation from the scenario of universality of the Yukawa couplings in both quark and lepton 
sectors. This would imply some moderate tuning among the Yukawa couplings. However, such a situation 
is considerably better compared to that of the SM, where a significant Yukawa parameter tuning is required.

In addition, the Cabbibo mixing together with the quark mixing in the $13$ and $23$ planes
are generated at one-loop level too. For this to happen, the $\mathbb{Z}_{2}^{(2)}$ symmetry has to be 
softly broken in the scalar sector, which is achieved by the trilinear 
$f_{234}\chi_{2}\chi_{3}\chi_{4}$ interaction term in the scalar potential (see Sec.~\ref{Sec:scalar-sector}). 

It is worth mentioning, as follows from Eqs.~(\ref{Lq}) and (\ref{Ll}), that the Yukawa interactions and 
mass terms are given in terms of the following parameters:
$y_{3}^{(Q)}$, $y_{nm}^{(Q)}$, $y_{n3}^{(Q)}$, 
$(x_{\chi})_{\alpha \beta}$, $(x_{\Psi})_{\alpha n}$, 
$(m_{\Psi})_{nm}$ ($n,m=1,2$).
Assuming all these parameters to be real except two of them amounts for a total of $28$ parameters, 
from which $8$ correspond to the quark Yukawa couplings, $16$ -- to the lepton Yukawa couplings, 
and $4$ -- to exotic lepton mass parameters. Apart from these, there are $8$ additional parameters 
useful to fit the SM fermion masses and mixing angles, which correspond to the scalar boson mass terms,
$m_{(\chi_{2}^*)_{23}}$, $m_{(\chi_{2}^*)_{32}}$, $m_{(\chi_{2}^*)_{13}}$, $m_{(\chi_{2}^*)_{31}}$, 
$m_{(\chi_{3}^*)_{32}}$, $m_{(\chi_{1})_{23}}$, $m_{(\chi_{1})_{32}}$ and $m_{(\chi_{4})_{1}}$ 
for the scalars in the internal lines of the Feynman diagrams in Fig.~\ref{Loopdiagramsq} and 
in the first diagram in Fig.~\ref{Loopdiagramsnu}. 

This gives a total of 36 input parameters. Of course, with this parametric freedom one can easily   
accommodate the experimental values for the 10 and 9 physical observables in the quark and lepton sectors, respectively. 
%
In this respect it is worth recalling again, that we do not pretend to predict the values of these observables, but the existing hierarchy between them.
The assumption about real Yukawa couplings, we made, corresponds to a particular benchmark scenario of CP-conserving Yukawa sector we chose here to simplify our discussion. Our model does not have a particular symmetry that forces some or all Yukawa couplings to be real hence enabling CP violation in our model (while indeed some of the Yukawa couplings can always be made real by phase redefinition of the fermionic fields). If the Yukawa couplings were allowed to be complex, 
the Yukawa sector would have a total of 60 parameters (not accounting for phase redefinitions of the fields), significantly increasing the parametric freedom compared to the considered CP-conserving benchmark scenario and hence not affecting our current predictions on SM fermion 
masses and mixings obtained here in this more constrained case.

The following comments about radiative stability of the symmetry breaking scheme (\ref{SB}) and the corresponding VEV pattern (\ref{VEVsinglets}) are in order. We note that this pattern is not protected from large radiative corrections by some symmetry. Therefore, to stabilize it we need to apply certain tuning of the model parameters. The corresponding conditions of vacuum stability are derived from  
the Coleman-Weinberg type 1-loop effective potential.
This analysis is beyond the scope of the present paper.  
However, since in our model the VEV hierarchy (\ref{VEVsinglets}) is rather moderate, not exceeding three orders of magnitude, we expect that 
the quadratic divergences -- dangerous for a strong hierarchy-- can be tamed here by a moderate tuning of the model parameters. At the same time, for the scales above $ v _ {\chi} $ in (\ref{VEVsinglets}),  where this is not possible, we assume that our model is embedded into a more fundamental setup with additional symmetries protecting the hierarchy up to the Planck scale.
The well-known examples of such setups are supersymmetry and warped five-dimensions.

\begin{table}[th]
\begin{tabular}{|c|c|c|c|c|}
\hline
& $\chi_{1}$ & $\chi_{2}$ & $\chi_{3}$ & $\chi_{4}$ \\ \hline
$\SU{3}{C}$ & $\mathbf{1}$ & $\mathbf{1}$ & $\mathbf{1}$ & $\mathbf{1}$ \\ 
\hline
$\SU{3}{L}$ & $\mathbf{3}$ & $\mathbf{3}$ & $\mathbf{3}$ & $\mathbf{3}$ \\ 
\hline
$\SU{3}{R}$ & $\overline{\mathbf{3}}$ & $\overline{\mathbf{3}}$ & $\mathbf{3}
$ & $\mathbf{1}$ \\ \hline
$\U{X}$ & $0$ & $0$ & $\frac{1}{3}$ & $-\frac{1}{3}$ \\ \hline
$\mathbb{Z}_{2}^{(1)}$ & $-1$ & $1$ & $-1$ & $-1$ \\ \hline
$\mathbb{Z}_{2}^{(2)}$ & $1$ & $1$ & $-1$ & $1$ \\ \hline
\end{tabular}%
\caption{Scalar assignments under $\SU{3}{C}\times \SU{3}{L}\times
\SU{3}{R}\times \U{X}\times \mathbb{Z}_{2}^{(1)}\times \mathbb{Z}_{2}^{(2)}$ symmetry.}
\label{tab:scalars}
\end{table}
\begin{table}[th]
\begin{tabular}{|c|c|c|c|c|c|c|c|}
\hline
& $Q_{nL}$ & $Q_{3L}$ & $Q_{nR}$ & $Q_{3R}$ & $L_{\alpha L}$ & $L_{\alpha R}$ & $\Psi_{nR}$ \\ \hline
$\SU{3}{C}$ & $\mathbf{3}$ & $\mathbf{3}$ & $\mathbf{3}$ & $\mathbf{3}$ & $%
\mathbf{1}$ & $\mathbf{1}$ & $\mathbf{1}$ \\ \hline
$\SU{3}{L}$ & $\overline{\mathbf{3}}$ & $\mathbf{3}$ & $\mathbf{1}$ & $%
\mathbf{1}$ & $\mathbf{3}$ & $\mathbf{1}$ & $\mathbf{1}$ \\ \hline
$\SU{3}{R}$ & $\mathbf{1}$ & $\mathbf{1}$ & $\overline{\mathbf{3}}$ & $%
\mathbf{3}$ & $\mathbf{1}$ & $\mathbf{3}$ & $\mathbf{1}$ \\ \hline
$\U{X}$ & $0$ & $\frac{1}{3}$ & $0$ & $\frac{1}{3}$ & $-\frac{1}{3}$ & $-%
\frac{1}{3}$ & $0$ \\ \hline
$\mathbb{Z}_{2}^{(1)}$ & $1$ & $1$ & $1$ & $-1$ & $1$ & $1$ & $-1$ \\ 
\hline
$\mathbb{Z}_{2}^{(2)}$ & $1$ & $-1$ & $1$ & $-1$ & $1$ & $1$ & $1$ \\ 
\hline
\end{tabular}
\caption{Fermion assignments under $\SU{3}{C}\times \SU{3}{L}\times
\SU{3}{R}\times \U{X}\times \mathbb{Z}_{2}^{(1)}\times \mathbb{Z}_{2}^{(2)}$. 
Here $n=1,2$ and $\protect\alpha = 1,2,3$.}
\label{tab:fermions}
\end{table}


\section{The neutrino sector} 
\label{Sec:neutrino}

\begin{figure}
    \centering
    \includegraphics[width = 0.9\textwidth]{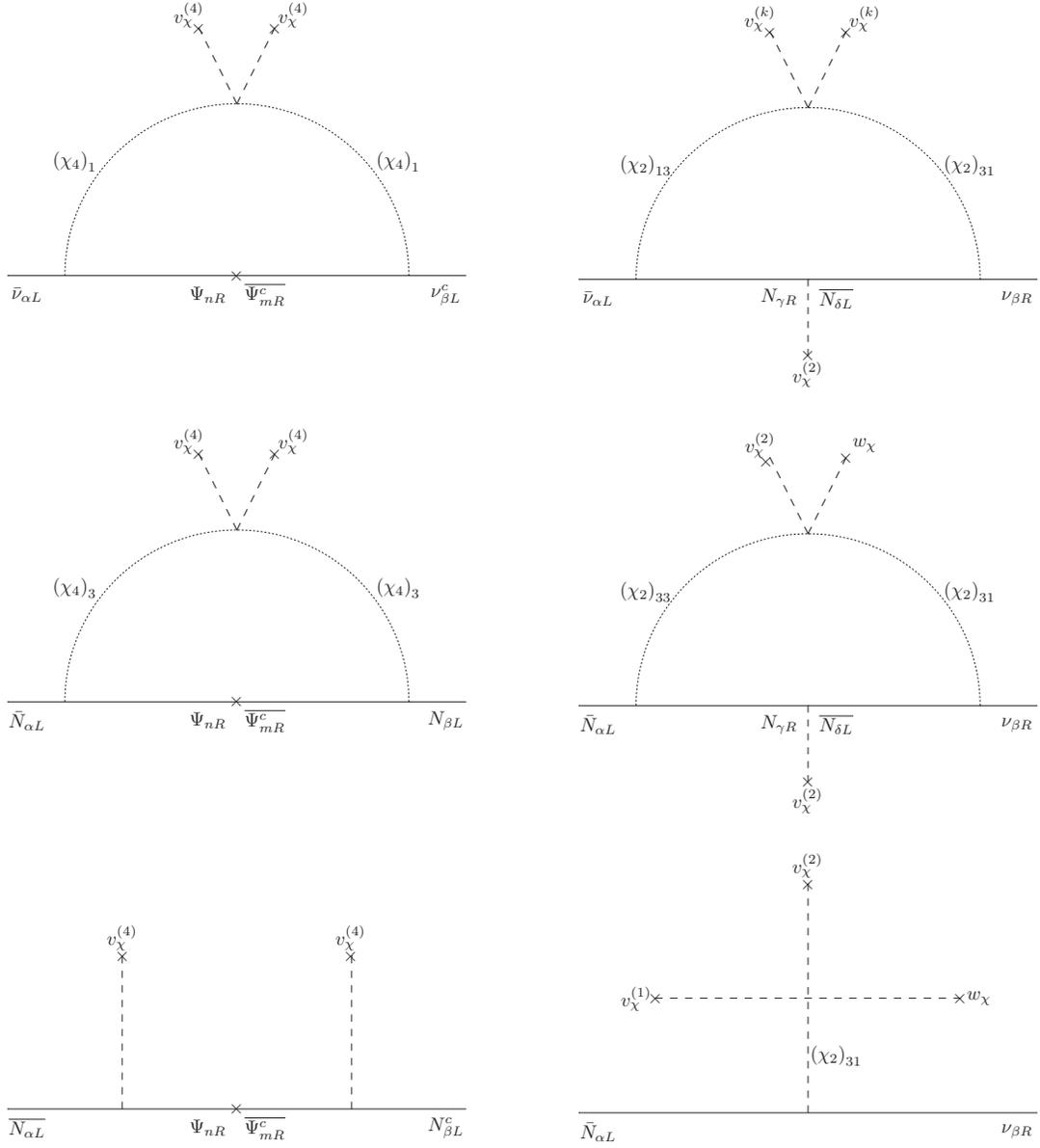}
    \caption{Feynman diagrams contributing to the entries of the neutrino mass
matrix in Eq.~\ref{Mnu}. Here, $n,m,k=1,2$ and $\protect\alpha ,\protect\beta ,
\protect\gamma =1,2,3$.}
\label{Loopdiagramsnu}
\end{figure}

From the neutrino Yukawa interactions (\ref{Ll}), we obtain the following mass terms:
\begin{equation}
-\mathcal{L}_{\rm mass}^{\left( \nu \right) } = 
\frac{1}{2}\overline{\Delta^{c}}^{T}M_{\nu }\Delta + {\rm h.c.} \,, 
\label{Lnu}
\end{equation}
where the neutrino basis and neutrino mass matrix (at one-loop order) are
given by, respectively,
\begin{eqnarray}
\Delta &=&\left( 
\begin{array}{c}
\nu_{\alpha L}^{c} \\ 
\nu_{\alpha R} \\ 
N_{\alpha L}^{c} \\ 
N_{\alpha R} \\ 
\Psi_{nR}
\end{array}
\right) \,,
\hspace{1cm}\hspace{1cm} \\
M_{\nu } &=&\left( 
\begin{array}{cccccc}
& \nu_{\beta L}^{c} & \nu_{\beta R} & N_{\beta L}^{c} & N_{\beta R} & \Psi_{mR} \\ \hline
\overline{\nu }_{\alpha L}| & M_{\overline{\nu }_{\alpha L}\nu _{\beta
L}^{c}} & M_{\overline{\nu }_{\alpha L}\nu _{\beta R}} & 0 & 0 & 0 \\ 
\overline{\nu _{\alpha R}^{c}}| & M_{\overline{\nu }_{\alpha L}\nu _{\beta
R}}^{T} & M_{\overline{\nu _{\alpha R}^{c}}\nu _{\beta R}} & 
M_{\overline{\nu_{\alpha R}^{c}}N_{\beta L}^{c}} & 0 & 0 \\ 
\overline{N}_{\alpha L}| & 0 & M_{\overline{\nu _{\alpha R}^{c}}N_{\beta
L}^{c}}^{T} & M_{\overline{N}_{\alpha L}N_{\beta L}^{c}} & \frac{v_{\chi
}^{\left( 2\right) }}{\sqrt{2}}\left( x_{\chi }\right) _{\alpha \beta } & 
\frac{v_{\chi }^{\left( 4\right) }}{\sqrt{2}}\left( x_{\Psi }\right)_{\alpha m} \\ 
\overline{N_{\alpha R}^{c}}| & 0 & 0 & \frac{v_{\chi }^{\left( 2\right) }}{
\sqrt{2}}\left( x_{\chi }\right) _{\beta \alpha } & 0 & 0 \\ 
\overline{\Psi _{nR}^{c}}| & 0 & 0 & \frac{v_{\chi }^{\left( 4\right) }}{
\sqrt{2}}\left( x_{\Psi }\right) _{n\beta } & 0 & \left( m_{\Psi }\right)_{nm}
\end{array}
\right) \,,
\label{Mnu}
\end{eqnarray}
with some of the sub-matrices generated at one-loop level from the Feynman diagrams 
in Fig.~\ref{Loopdiagramsnu}. The sub-matrices appearing in Eq.~(\ref{Mnu}) are given by:
\begin{eqnarray}
M_{\overline{\nu }_{\alpha L}\nu _{\beta L}^{c}} &=&\sum_{n=1}^{2}\frac{
\left( x_{\Psi }\right) _{\alpha n}\left( x_{\Psi }\right) _{\beta n}m_{\Psi
_{n}}}{16\pi ^{2}}f\left( m_{\Psi _{n}},m_{\func{Re}\left( \chi _{4}\right)
_{1}}^{2},m_{\func{Im}\left( \chi _{4}\right) _{1}}^{2}\right) ,\hspace{1cm}
\alpha ,\beta =1,2,3, \\
M_{\overline{\nu }_{\alpha L}\nu _{\beta R}} &=&\sum_{\gamma =1}^{3}\frac{
\left( x_{\chi }\right) _{\alpha \gamma }\left( x_{\chi }\right) _{\beta
\gamma }m_{N_{\gamma }}}{16\pi ^{2}}f\left( m_{N_{\gamma }},m_{\func{Re}
\left( \chi _{2}\right) _{13}}^{2},m_{\func{Im}\left( \chi _{2}\right)
_{13}}^{2}\right) ,\hspace{1cm}m_{N_{\gamma }}=\left( x_{\chi }\right)
_{\gamma }\frac{v_{\chi }}{\sqrt{2}} \,, \\
M_{\overline{\nu _{\alpha R}^{c}}N_{\beta L}^{c}} &=&\frac{\kappa
_{1122}^{\left( \chi \right) }v_{\chi }^{\left( 1\right) }v_{\chi }^{\left(
2\right) }w_{\chi }}{2\sqrt{2}m_{\left( \chi _{2}\right) _{31}}^{2}}\left(
x_{\chi }\right) _{\alpha \beta } \,, \\
M_{\overline{\nu _{\alpha R}^{c}}\nu _{\beta R}} &=&\left( \frac{\kappa
_{1122}^{\left( \chi \right) }v_{\chi }^{\left( 1\right) }v_{\chi }^{\left(
2\right) }w_{\chi }}{2\sqrt{2}m_{\left( \chi _{2}\right) _{31}}^{2}}\right)
^{2}\left( x_{\chi }\right) _{\alpha \gamma }M_{\overline{N}_{\gamma
L}N_{\delta L}^{c}}^{-1}\left( x_{\chi }\right) _{\delta \beta }, \\
M_{\overline{N}_{\alpha L}N_{\beta L}^{c}} &=&\frac{\left( v_{\chi }^{\left(
4\right) }\right) ^{2}}{2}\left( x_{\Psi }\right) _{\alpha n}\left( m_{\Psi
}^{-1}\right) _{nm}\left( x_{\Psi }\right) _{m\beta } \,, \\
f\left( m_{F},m_{R},m_{I}\right) &=&\left[ \frac{m_{R}^{2}}{%
m_{R}^{2}-m_{F}^{2}}\ln \left( \frac{m_{R}^{2}}{m_{F}^{2}}\right) -\frac{
m_{I}^{2}}{m_{I}^{2}-m_{F}^{2}}\ln \left( \frac{m_{I}^{2}}{m_{F}^{2}}\right)
\right] \,.
\end{eqnarray}
where $\alpha =1,2,3$ and $n=1,2$. In addition, the entries denoted by $X$
and $y$ are generated at tree- and one-loop levels, respectively. 

The light active neutrino masses arise from a combination of radiative and type-I
seesaw mechanisms (with the Dirac neutrino mass matrix generated at one-loop
order). This mechanism in a more general setup has been recently proposed in Ref. \cite{Arbelaez:2019wyz}. Thus, the mass matrix for the light neutrinos takes the form
\begin{equation}
\widetilde{M}_{\nu } = M_{\overline{\nu }_{L}^{T}\nu _{L}^{c}} - AM_{S}^{-1}A^{T} \,,
\end{equation}
where the matrices $M_{\overline{\nu }_{L}^{T}\nu _{L}^{c}}$ and $A$ are
generated at one-loop level whereas $M_{S}$ receives tree-level and one-loop
contributions. The matrices $A$ and $M_{S}$ are found as follows
\begin{eqnarray}\label{eq:nu-mass-1}
A=\left( 
\begin{array}{c}
M_{\overline{\nu }_{\alpha L}\nu_{\beta R}}^{T} \\ 
0 \\ 
0 \\ 
0
\end{array}
\right)^{T} \,,
\hspace{1cm}\hspace{1cm}M_{S}=\left( 
\begin{array}{cccc}
M_{\overline{\nu _{\alpha R}^{c}}\nu _{\beta R}} & M_{\overline{\nu _{\alpha
R}^{c}}N_{\beta L}^{c}} & 0 & 0 \\ 
M_{\overline{\nu _{\alpha R}^{c}}N_{\beta L}^{c}}^{T} & M_{\overline{N}
_{\alpha L}N_{\beta L}^{c}} & \frac{v_{\chi }^{\left( 2\right) }}{\sqrt{2}}
\left( x_{\chi }\right) _{\alpha \beta } & \frac{v_{\chi }^{\left( 4\right) }
}{\sqrt{2}}\left( x_{\Psi }\right) _{\alpha m} \\ 
0 & \frac{v_{\chi }^{\left( 2\right) }}{\sqrt{2}}\left( x_{\chi }\right)
_{\beta \alpha } & 0 & 0 \\ 
0 & \frac{v_{\chi }^{\left( 4\right) }}{\sqrt{2}}\left( x_{\Psi }\right)
_{n\beta } & 0 & \left( m_{\Psi }\right) _{nm}
\end{array}
\right) \,.
\end{eqnarray}
Considering $v\ll w_{\chi }$ $\ll m_{\left( \chi _{2}\right)_{31}}^{2}\sim
v_{\chi }^{\left( 1\right) }\sim v_{\chi }^{\left( 2\right) } \ll m_{\Psi }$,
from the previous relations it follows that the entries of the light active
neutrino mass matrix can be approximated as
\begin{eqnarray}\label{eq:numass-2}
\left( \widetilde{M}_{\nu }\right) _{\alpha \beta }\simeq M_{\overline{\nu }
_{\alpha L}\nu _{\beta L}^{c}}=\sum_{n=1}^{2}\frac{\left( x_{\Psi }\right)
_{\alpha n}\left( x_{\Psi }\right) _{\beta n}m_{\Psi _{n}}}{16\pi ^{2}}
f\left( m_{\Psi _{n}},m_{\func{Re}\left( \chi _{4}\right) _{1}}^{2},m_{\func{
Im}\left( \chi _{4}\right) _{1}}^{2}\right) \,.
\end{eqnarray}
In order to get an approximate expression for the physical sterile neutrino
mass matrices, we assume $w_{\chi }\sim \mathcal{O}(1)$ TeV, $m_{\Psi }\sim 
\mathcal{O}(10^{3})$ TeV and consider $v_{\chi }^{\left( 1\right) }\sim
v_{\chi }^{\left( 2\right) }\sim v_{\chi }^{\left( 4\right) }\sim \mathcal{O}
(10^{2})$ TeV. Taking the Yukawa couplings of order unity, we thus recover the
following estimates:
\begin{equation*}
M_{\overline{\nu_{\alpha R}^{c}}N_{\beta L}^{c}}\sim \mathcal{O}(1)\hspace{1mm}\mathrm{TeV} \,,
\hspace{1cm}\hspace{1cm}
M_{\overline{N}_{\alpha L}N_{\beta
L}^{c}}\sim \mathcal{O}(10)\hspace{1mm}\mathrm{TeV} \,.
\end{equation*}
In the limit when the remaining sterile neutrinos are very heavy, i.e. when their mass
matrices approach $\pm x_{\chi }v_{\chi }^{(2)}/\sqrt{2}$ and $m_{\Psi}$,
the lightest physical sterile neutrino mass matrix can be
approximated as $M_{\overline{\nu _{\alpha R}^{c}}\nu_{\beta R}}$.

Assuming $w_\chi \sim v_\chi^{(2)} \sim \mathcal{O}(10) \mathrm{TeV}$, $v_\chi^{(1)} \sim v_\chi^{(4)}\sim \mathcal{O}(10^3) \mathrm{TeV}$, 
$m_{\Psi}\sim \mathcal{O}(10^4) \mathrm{TeV}$, and $x_\Psi\sim 10 x_\chi \sim 1$, one can roughly estimate 
	\begin{equation*}
	M_{\overline{\nu_R^c} N_R} \sim \mathcal{O}(1) \, \mathrm{TeV}\,, \hs 
	M_{\overline{\nu_R^c} \nu_R} \sim \mathcal{O}(10)\, \mathrm{TeV} \,, \hs 
	M_{\overline{N_R^c} N_R} \sim \mathcal{O}(100)\, \mathrm{TeV} \,, \hs  
	\frac{x_\chi v_\chi^{(2)}}{\sqrt{2}}\sim \mathcal{O}(1)\, \mathrm{TeV} \,. 
	\end{equation*}
In this case, the lightest sterile neutrino is identified with the right handed neutral lepton $N_{aR}$ and their mixing 
mass matrix is proportional to $\frac{x_\chi v_\chi^{(2)}}{\sqrt{2}}$.

\section{Implications for quarks and charged leptons} 
\label{Sec:SM-fermions}

Expanding the Lagrangian in Eq.~\eqref{Lq} and considering, for simplicity, that
\be
y^{(Q)}_{nm} \equiv y\qquad\text{for}~m=n
\ee
and
\be
y^{(Q)}_{nm} \equiv h\qquad\text{for}~m \neq n \,,
\ee
one obtains five massless quarks, corresponding to all down-type quarks as well as the first and second generation up-type ones. The top quark mass is readily generated at tree level,
\be
m_t^2 = \frac12 y^2_3 v^2 \frac{v_\chi
	^2}{v_\chi^2 + \omega_\chi^2}\,.
\ee
The model also predicts three heavy vector-like quarks with the following masses
\be
m^2_{V_1} \simeq \frac12 y_3^2 (v_\chi^2 + \omega_\chi^2) \,,
\qquad
m^2_{V_2} = \frac12 v_\chi^2 \(h+y\)^2  \,,
\qquad
m^2_{V_3} = \frac12 v_\chi^2 \(h-y\)^2\,.
\ee
Since the top Yukawa coupling $y_3$ is large, $V_1$ will always be of order $100~\mathrm{TeV}$ and can
only be probed at a future FCC facility. However, if the couplings $h$ and $y$ are somewhat smaller, say
of order $0.01$, then both $V_2$ and $V_3$ can be at a TeV scale and hence at the reach of the LHC. In a
third scenario, if both $h$ and $y$ are of order one and not far off each other, then only $V_3$ can
become light enough to be probed at the LHC. Furthermore, as follows from Eq.~\eqref{Lq}, the exotic
quarks can decay into a SM quark and either a neutral or a charged scalar and can be pair-produced 
at the LHC via Drell-Yan and gluon fusion processes, mediated by charged gauge bosons and gluons, 
respectively. A detailed study of the collider phenomenology of the model is beyond the scope of this
paper and is left for future dedicated explorations. 

For the charged lepton sector we follow a similar 
approach, namely, we expand the first term in Eq.~\eqref{Ll} and then set
\begin{eqnarray}
&& \(x_\chi\)_{\alpha \beta} = x\qquad\text{for}~\alpha=\beta \,, \\
&& \(x_\chi\)_{\alpha \beta} = k\qquad\text{for}~\alpha \neq \beta\,.
\end{eqnarray}
The model also contains three heavy vector-like leptons with the following mass spectrum:
\be
m^2_{E_1} = \frac12 v_\chi^2 \(2 k + x\)^2 \,,
\qquad
m^2_{E_2} = m^2_{E_3} = \frac12 v_\chi^2 \(k - x\)^2\,.
\ee
Similarly to the quark sector, if both $k$ and $x$ are of order $0.01$ or slightly smaller, then the 
model predicts three generations of vector-like leptons at the TeV scale or even below. On the other 
hand, if such Yukawa couplings are of order one, then only $E_2$ and $E_3$ can become sufficiently 
light to be at the reach of the LHC measurements.

Note, the above assumptions on the parameters of the Yukawa sector are made to keep the discussion as simple 
as possible and correspond to a particular benchmark scenario chosen for simplicity. We stress that our model predicts 
that the SM charged leptons, neutrinos and quarks (except for the top quark) are massless at tree level. This is a generic feature 
of the considered model, independent of the assumptions made about the parameters of the SM fermion mass matrices.

\section{The scalar sector}
\label{Sec:scalar-sector}

We start our analysis of the scalar sector by considering the first breaking step in the chain \eqref{SB}.
The scalar potential of the low-scale trinification theory, which is invariant under the transformations
specified in Tab.~\ref{tab:scalars}, reads
\be
\bal
V\(\chi_a\) =&  \dfrac12 \sum_{a = 1}^{2} \mu _{\left( \chi \right) a}^{2} \Big(\chi_a\Big)^r_l \Big(\chi_a^\dagger\Big)^l_r 
+
\dfrac12 \mu _{\left( \chi \right) 3}^{2}  \Big(\chi_3\Big)_{r\,l} \Big(\chi_3^\dagger\Big)^{l\,r} 
+
\dfrac12 \mu _{\left( \chi \right) 4}^{2} \Big(\chi_4\Big)_{l} \Big(\chi_4^\dagger\Big)^{l}
\\
&
+ \dfrac14 \sum_{b \geq a}^{2} \kappa_{ab}^{\left( \chi \right) }  \Big(\chi_a\Big)^r_l \Big(\chi_a^\dagger\Big)^l_r \Big(\chi_b\Big)^{r'}_{l'} \Big(\chi_b^\dagger\Big)^{l'}_{r'}
+
\dfrac{1}{4} \kappa_{33}^{\left( \chi \right) }  \Big(\chi_3\Big)_{r\,l} \Big(\chi_3^\dagger\Big)^{l\,r} \Big(\chi_3\Big)_{r'\,l'} \Big(\chi_3^\dagger\Big)^{l'\,r'}
\\
&
+
\dfrac14 \sum_{a = 1}^{2} \kappa_{a3}^{\left( \chi \right) }  \Big(\chi_a\Big)^{r}_{l} 
\Big(\chi_3\Big)_{r'\,l'} 
\Big(\chi_3^\dagger\Big)^{l'\,r'}
\Big(\chi_a^\dagger\Big)^{l}_{r} 
+
\dfrac14 \sum_{a = 1}^{2}
\kappa_{a4}^{\left( \chi \right) }  \Big(\chi_a\Big)^r_l \Big(\chi_a^\dagger\Big)^l_r \Big(\chi_4\Big)_{l'} \Big(\chi_4^\dagger\Big)^{l'}
\\
&
+
\dfrac14
\kappa_{34}^{\left( \chi \right) }  \Big(\chi_3\Big)_{r\,l} \Big(\chi_3^\dagger\Big)^{l\,r} \Big(\chi_4\Big)_{l'} \Big(\chi_4^\dagger\Big)^{l'}
+
\dfrac14
\kappa_{44}^{\left( \chi \right) }  \Big(\chi_4\Big)_{l} \Big(\chi_4^\dagger\Big)^{l} \Big(\chi_4\Big)_{l'} \Big(\chi_4^\dagger\Big)^{l'}
\\
&
+ \dfrac14 \sum_{b \geq a}^{2} \kappa_{ab}^{\prime \left( \chi \right) } \Big(\chi_a^\dagger\Big)^{l'}_r  
\Big(\chi_a\Big)^r_l 
\Big(\chi_b^\dagger\Big)^{l}_{r'}
\Big(\chi_b\Big)^{r'}_{l'} 
+
\dfrac{1}{4} \kappa_{33}^{\prime \left( \chi \right) }  \Big(\chi_3\Big)_{r\,l} \Big(\chi_3^\dagger\Big)^{l\,r} \Big(\chi_3\Big)_{r'\,l'} \Big(\chi_3^\dagger\Big)^{l'\,r'}
\\
&
+ \kappa_{12}^{\prime \prime \left( \chi \right) }  \Big(\chi_1\Big)^r_l \Big(\chi_1^\dagger\Big)^{l}_{r'} \Big(\chi_2\Big)^{r'}_{l'} \Big(\chi_2^\dagger\Big)^{l'}_{r}
+\kappa_{12}^{\prime \prime \prime \left( \chi \right)}  \Big(\chi_1\Big)^r_l \Big(\chi_1^\dagger\Big)^{l'}_{r'} \Big(\chi_2\Big)^{r'}_{l'} \Big(\chi_2^\dagger\Big)^{l}_{r}  
\\
&
+
\dfrac14 \sum_{a = 1}^{2}
\left[ \kappa_{a3}^{\prime \left( \chi \right) } 
\Big(\chi_a\Big)^r_l 
\Big(\chi_a^\dagger\Big)^{l'}_r 
\Big(\chi_3\Big)_{r'\,l'} 
\Big(\chi_3^\dagger\Big)^{l\,r'}
\right.
\\
&
\left.
+ \kappa_{a3}^{\prime \prime \left( \chi \right) }  
\Big(\chi_a\Big)^r_l 
\Big(\chi_a^\dagger\Big)^{l}_{r'} 
\Big(\chi_3\Big)_{r\,l'} 
\Big(\chi_3^\dagger\Big)^{l'\,r'}
+\kappa_{a3}^{\prime \prime \prime \left( \chi \right)}  \Big(\chi_a\Big)^r_l 
\Big(\chi_a^\dagger\Big)^{l'}_{r'} 
\Big(\chi_3\Big)_{r\,l'} 
\Big(\chi_3^\dagger\Big)^{l\,r'}  
\right]
\\
&
+\dfrac14 \sum_{a = 1}^{2}
\kappa_{a4}^{\prime \left( \chi \right) }  \Big(\chi_a\Big)^r_l \Big(\chi_a^\dagger\Big)^{l'}_r \Big(\chi_4\Big)_{l'} \Big(\chi_4^\dagger\Big)^{l}
+
\dfrac14 
\kappa_{34}^{\prime \left( \chi \right) }  \Big(\chi_3\Big)_{r\,l} \Big(\chi_3^\dagger\Big)^{l'\,r} \Big(\chi_4\Big)_{l'} \Big(\chi_4^\dagger\Big)^{l}
\\
&
+
\left\{\epsilon^{l l' l''} \epsilon_{r r' r''} \left[f_2 \Big(\chi_2\Big)^r_l \Big(\chi_2\Big)^{r'}_{l'} \Big( \chi_2\Big)^{r''}_{l''}
+
f_{12} \Big(\chi_1\Big)^r_l \Big(\chi_1\Big)^{r'}_{l'} \Big( \chi_2\Big)^{r''}_{l''} \right]
+
\rm{h.c.}
\right\} \,,
\label{Vtri}
\eal
\ee
where $l$ and $r$ denote $\SU{3}{L}$ and $\SU{3}{R}$ indices, respectively, while $a$ is a scalar flavour index. Fundamental and anti-fundamental $\SU{3}{L} \times \SU{3}{R}$ indices are written in superscript and subscript, respectively. Note that the potential $V\(\chi_a\)$, in the limit $f_2 \to 0$, is invariant under an accidental global $\U{acc}$ phase rotation, which can be defined as
\be
\chi_{1,2} \to e^{-i q_{1,2}\theta} \chi_{1,2}^\prime\,,
\label{acc-1}
\ee
and where the global charges of the $\chi_1$ and $\chi_2$ fields can be defined as $q_1 = \tfrac{1}{2}$ and $q_2 = -1$, respectively. Therefore, in the limit $f_2 \to 0$, the vacuum of the theory, after the breaking $\SU{3}{L} \times \SU{3}{R} \times \U{X} \times \U{acc} \to \SU{2}{L} \times \SU{2}{R} \times \U{B-L}$, features 11 Goldstone bosons, where one of them becomes physical since it corresponds to the breaking of the global $\U{acc}$ generator. In particular, such a Goldstone boson, which we will denote as $A$ in what follows, is a CP-odd scalar resulting from a combination of the imaginary parts of the $\(\chi_{1,2}\)^3_3$ components. Furthermore, while $f_2 \neq 0$ violates $\U{acc}$, contractions with the Levi-Civita symbols in both $f_2$ and $f_{12}$ terms imply that the $\(\chi_{1,2}\)^3_3$ components are still protected from acquiring mass. 

On the another hand, as we already mentioned in Sec. \ref{Sec:Model}, a complete description of quark mixing and, in particular, small 
%
CKM matrix elements in $13$ and $23$ planes requires a small cubic interaction of the form $f_{234} \chi_2 \chi_3 \chi_4$ softly breaking 
$\mathbb{Z}_2^{(2)}$. This symmetry breaking effect is transmitted by radiative corrections 
to the Yukawa sector. 
%
The $\mathbb{Z}_2^{(2)}$, being  softly broken by this trilinear term, still protects the smallness of  the parameter $f_{234}$ and its effect in all the sectors of the model from large radiative corrections.

Interestingly, 
by introducing a small explicit violation of both $\mathbb{Z}_2^{(1)}$ and $\mathbb{Z}_2^{(2)}$ with the following soft-breaking terms
\be
\bal
V^{\slashed{\mathbb{Z}}_2}_\mathrm{soft} = 
\mu _{\left( \chi \right) 12}^{2} \Big(\chi_1\Big)^r_l \Big(\chi_2^\dagger\Big)^l_r
+ f_{234} \epsilon^{l l' l''} \Big( \chi_2\Big)^{r}_{l} \Big(\chi_3\Big)_{r\,l'} \Big(\chi_4\Big)_{l''}
+ \rm{h.c.}\,,
\label{Vsoft}
\eal
\ee
we are not only allowing for the generation of small entries in the CKM matrix, but also softly breaking $\U{acc}$ by means of $\mu_{(\chi)12}^{2}$ term. The latter promotes the CP-odd scalar to a pseudo-Goldstone boson with mass
\be
m_A^2 = - 2 \mu_{(\chi)12}^2\,.
\label{mu12}
\ee
typically referred to as Majoron. This can be understood from a comparison with the conventional Majoron models with type-I seesaw mechanism \cite{Gelmini:1984pe,Berezinsky:1993fm,Gu:2010ys,Lattanzi:2013uza,Garcia-Cely:2017oco,Heeck:2017wgr,Brune:2018sab,Heeck:2019guh,Abe:2020dut}. There, a complex SM-singlet scalar couples directly to active Majorana neutrinos. However, in our model, in place of a complex singlet we have a bi-triplet, $\chi_2$, where the Majoron $A$, belonging to the $\(\chi_2\)^3_3$ component, only couples to the sterile neutrinos in the third entry of both $L_\mathrm{L,R}$. In turn, this means that tree-level couplings to the EW gauge bosons are always suppressed by a tiny mixing with active neutrinos suppressing the loop-induced Majoron decays into photons, $A\to \gamma \gamma$. This implies that a trinification Majoron can become a light DM candidate if its mass is below a MeV scale. This will further be discussed in Sec.~\ref{Sec:DM}.
Note that due to unbroken CP-symmetry in the scalar sector, this particle only forms quadratic and quartic interactions in the scalar potential of the low-energy effective theory. It is also worth mentioning that \eqref{Vsoft} prevents the formation of domain walls in the early Universe which would in principle appear from the spontaneous breaking of $\mathbb{Z}_2^{(1)}$.

For a cleaner analysis of the mass spectrum, we will make a few simplifying assumptions on the theory parameters. First, inspired by the gauge quantum numbers, we will consider that the couplings involving $\chi_1$ and $\chi_2$ are identical, i.e.
\be
\bal
&\mu _{\left( \chi \right) 1}^{2} = \mu _{\left( \chi \right) 2}^{2} \gg \mu _{\left( \chi \right) 12}^{2}
\,,
~
\kappa_{11}^{\left( \chi \right) } = \kappa_{22}^{\left( \chi \right) } \equiv \kappa_{1}^{\left( \chi \right) }
\,,
~
\kappa_{11}^{\prime\left( \chi \right) } = \kappa_{22}^{\prime\left( \chi \right) } \equiv \kappa_{1}^{\prime\left( \chi \right) }
\,,
~
\kappa_{13}^{\left( \chi \right) } = \kappa_{23}^{\left( \chi \right) }
\,,
~
\\
&
\kappa_{12}^{\left( \chi \right) } = \kappa_{12}^{\prime\,\left( \chi \right) } = \kappa_{12}^{\prime \prime \,\left( \chi \right) } = \kappa_{12}^{\prime \prime \prime\,\left( \chi \right) }
\,,
~
\kappa_{13}^{\prime\,\left( \chi \right) } = \kappa_{13}^{\prime \prime \,\left( \chi \right) } = \kappa_{13}^{\prime \prime \prime\,\left( \chi \right) } =
\kappa_{23}^{\prime\,\left( \chi \right) } = \kappa_{23}^{\prime \prime \,\left( \chi \right) } = \kappa_{23}^{\prime \prime \prime\,\left( \chi \right) } \equiv
\kappa_{13}^{\left( \chi \right)}
\,,
\\
&
\kappa_{14}^{\left( \chi \right) } 
= \kappa_{24}^{\left( \chi \right) }
= \kappa_{14}^{\prime\left( \chi \right) } 
= \kappa_{24}^{\prime\left( \chi \right) }
\,,
~
f_1 = f_{12} \equiv f\,.
\eal
\label{simp}
\ee
Note that such conditions, together with the transformation properties in Tab.~\ref{tab:scalars}, imply that 
\be
v_{\chi }^{\left( 1\right) } =
v_{\chi }^{\left( 2\right) } \equiv v_{\chi }\,.
\ee
Solving the tadpole equations with respect to the $ v_{\chi }$ and $v_{\chi }^{\left( 4\right) }$ \vevs~we get
\be
\mu _{\( \chi \) 1}^{2} = -\dfrac12 \[v_{\chi }^2 \( \kappa_{1}^{\( \chi \)} + \kappa_{1}^{\prime\( \chi \)} + 2 \kappa_{12}^{\( \chi \)} \) + {v_{\chi }^{\( 4\) }}^2 \kappa_{14}^{\( \chi \)} + 4 \mu_{\left( \chi \right) 12}^{2}  \]\,, \qquad
\mu _{\( \chi \) 4}^{2} = -\dfrac12 \( 2 v_{\chi }^2 \kappa_{14}^{\( \chi \)} + {v_{\chi }^{\( 4\) }}^2 \kappa_{4}^{\( \chi \)} \)\,.
\ee
Before proceeding, and in order to understand how the gauge structure and scalar mixing splits the trinification scalar representations, let us first note that we can decompose the $\chi_a$ in a total of three $\SU{2}{R} \times \SU{2}{L}$ bi-doublets, three $\SU{2}{R}$ doublets, denoted as R-doublets in what follows, four $\SU{2}{L}$ doublets, which we will call L-doublets, as well as eight singlets corresponding to the $\(\chi_{1,2,3}\)^3_3$ and $\(\chi_4\)_3$ components. Provided that the Goldstone bosons correspond to one L-doublet, one R-doublet as well as two real singlets, the physical fields can be decomposed in three bi-doublet blocks, two R-doublet and three L-doublet blocks, and six singlets. Now, one should note that in the vacuum of the theory, the $f_{234}$ cubic coupling splits two of the bi-doublets into four L-doublets, while the single bi-doublet left in the scalar spectrum results from the fact that there are no \vevs~in $\chi_3$. For the same reason, out of the six singlets, four are real and two form a complex one charged under $\U{B-L}$. In summary, we can list the physical scalars after the breaking of the trinification symmetry as
\begin{itemize}
    \item 1 bi-doublet $\Sigma$\,,
    \item 7 L-doublets $\mathrm{L}_{1,\ldots,7}$\,,
    \item 2 R-doublets $\mathrm{R}_{1,2}$\,,
    \item 1 complex singlet $\sigma$\,,
    \item 3 real CP-even singlets $\varphi_{1,2,3}$\,,
    \item 1 Majoron $A$\,.
\end{itemize}

\subsubsection{A minimal light scalar sector}

To visualize the model's behaviour at low-energy scales it is instructive to look at a numerical example. Here, 
we will use our freedom to set numerical values that: 1) enable us to sufficiently split the mass spectrum in order 
to obtain a minimal viable low-energy effective theory, and 2) ensure that such a scenario is simple enough to clearly highlight the most important features of the model under consideration. First, we set the following scales
\be
\bal
&
v_{\chi } = 160~\mathrm{TeV} \,,
\qquad
v_{\chi }^{\( 4\) } = 150~\mathrm{TeV}\,,
\qquad
\mu _{\( \chi \) 3} = 150~\mathrm{TeV}\,,
\\
&
\mu_{\left( \chi \right) 12}^{2} = -\(2.3442~\mathrm{TeV}\)^2\,,
\qquad
f = 0.1~\mathrm{TeV}\,,
\qquad
f_{234} = 0.01~\mathrm{TeV}
\,.
\eal
\label{VEVs-num}
\ee
For the quartic couplings, in addition to the simplifying assumptions in Eq.~\eqref{simp}, we have also considered that quartic interactions involving one single scalar flavour provide the leading contributions and are of order $\mathcal{O}(1)$, while the remaining ones are below $\mathcal{O}(0.1)$. Such a behaviour can typically be explained with flavour symmetries engineered to forbid tree-level couplings between different representations of a UV complete theory. For our benchmark example, we choose for the quartics the following sizes:
\be
\bal
&
\kappa_{1}^{\( \chi \)} = 1.1 \,, \quad \kappa_{4}^{\( \chi \)} = 0.95\,, \quad
\kappa_{1}^{\prime\(\chi\)} = -2.0\times 10^{-3}\,,
\quad
\kappa_{12}^{\( \chi \)} = -6.5 \times 10^{-3} \,,
\\
&
\kappa_{13}^{\( \chi \)} = 8.0 \times 10^{-2}\,,
\quad
\kappa_{14}^{\( \chi \)} = -4.0 \times 10^{-2}\,,
\quad
\kappa_{34}^{\( \chi \)} = 4.9 \times 10^{-2}\,.
\eal
\label{quartics-num}
\ee
If we now choose a criterion to denote \textit{light states} whenever their mass is below the $10~\mathrm{TeV}$ threshold, the input values in Eqs.~\eqref{VEVs-num} and \eqref{quartics-num} result in a light L-doublet, a light R-doublet and a Majoron 
with masses
\be
m_{\mathrm{L}_1} \approx m_{\mathrm{R}_1} \approx 9.7~\mathrm{TeV}
\qquad
\text{and}
\qquad
m_\mathrm{A} \approx 3.3~\mathrm{TeV}\,,
\label{light}
\ee
respectively.
We also obtain five \textit{next-to-light} L-doublets, i.e.~with masses between $10$ and $20~\mathrm{TeV}$, whose values read
\be
m_{\mathrm{L}_2} \approx 14.2~\mathrm{TeV}\,,
\quad 
m_{\mathrm{L}_3} \approx 14.4~\mathrm{TeV}\,,
\quad
m_{\mathrm{L}_4} \approx 14.6~\mathrm{TeV}\,,
\quad
m_{\mathrm{L}_5} \approx 16.2~\mathrm{TeV}\,,
\quad
\text{and}
\quad
m_{\mathrm{L}_6} \approx 19.2~\mathrm{TeV}\,.
\ee
Finally, in a third category, we group those states that we denote as \textit{heavy} whose masses read
\be
\bal
m_{\mathrm{L}_7} &\approx 111~\mathrm{TeV}\,,
\qquad
m_{\mathrm{R}_2} \approx 112~\mathrm{TeV}\,,
\qquad
m_{\Sigma} \approx 109~\mathrm{TeV}\,,
\qquad
m_{\sigma} \approx 117~\mathrm{TeV}\,,
\\
m_{\varphi_1} &\approx 103~\mathrm{TeV}\,,
\qquad
m_{\varphi_2} \approx 118~\mathrm{TeV}\,,
\qquad
m_{\varphi_3} \approx 119~\mathrm{TeV}\,.
\eal
\ee 

Inspired by the numerical example above, we consider a minimal scenario for the $\SU{2}{L} \times \SU{2}{R} \times \U{B-L}$ theory, typically referred to as the LR symmetric theory, where the scalar content can be reduced to $\mathrm{L}_1$, $\mathrm{R}_1$ and $\mathrm{A}$. 

Note that other parameter choices may provide a low-energy limit with a richer $\SU{2}{L}$ L-doublet content. In what follows, we recast our L- and R-doublets as $\mathrm{L}$ and $\mathrm{R}$ respectively. The most generic renormalizable scalar potential can be written as
\be
\bal
V_\mathrm{LR} =& \mu_\mathrm{L}^2 \mathrm{L}^\dagger \mathrm{L}
+ \mu_{\mathrm{R}}^2 \mathrm{R}^\dagger \mathrm{R} 
+ \mu_\mathrm{A}^2 \mathrm{A}^2
+ \lambda_\mathrm{L} \abs{\mathrm{L}^\dagger \mathrm{L}}^2
+
\lambda_\mathrm{R} \abs{\mathrm{R}^\dagger \mathrm{R}}^2
+
\lambda_\mathrm{A} \mathrm{A}^4
\\
+
&
\lambda_{\mathrm{L} \mathrm{R}} \(\mathrm{L}^\dagger \mathrm{L}\) \(\mathrm{R}^\dagger \mathrm{R}\)
+ \lambda_{\mathrm{A} \mathrm{L}} \mathrm{A}^2 \(\mathrm{L}^\dagger \mathrm{L}\)
+
\lambda_{\mathrm{A} \mathrm{R}} \mathrm{A}^2 \(\mathrm{R}^\dagger \mathrm{R}\) \,.
\eal
\label{VLR}
\ee
It follows from Eq.~\eqref{VEVs-tri} that both the LR and the EW symmetries can be broken by the vacuum assignment
\begin{eqnarray}
\left\langle \mathrm{L} \right\rangle &=&\left( 
\begin{array}{c}
\frac{v}{\sqrt{2}}\\ 
0
\end{array}
\right) \,,\hspace{1cm}\hspace{1cm}
\left\langle \mathrm{R}\right\rangle =\left( 
\begin{array}{c}
\frac{\omega_\chi}{\sqrt{2}} \\ 
0
\end{array}
\right) \,, 
\label{LR-EW-VEVs}
\end{eqnarray}%
where the solutions of the tadpole equations are given by
\be
\mu_\mathrm{L}^2 = -\frac14\( 2 v^2 \lambda_\mathrm{L} + \omega_\chi^2 \lambda_{\mathrm{L} \mathrm{R}}\)\,,
\qquad
\mu_{\mathrm{R}}^2 = - \frac12 \(2 \omega_\chi^2 \lambda_{\mathrm{R}} + v^2 \lambda_{\mathrm{L} \mathrm{R}}\)\,.
\ee
The two neutral CP-even scalar masses read
\be
m^2_{H,h} = v^2 \lambda_\mathrm{L} + \omega^2 \lambda_{\mathrm{R}} \pm \sqrt{v^4 \lambda_\mathrm{L}^2 + v^2 \omega_\chi^2 (\lambda_{\mathrm{L} \mathrm{R}}^2 - 2 \lambda_{\mathrm{L}} \lambda_{\mathrm{R}}  ) + \omega_\chi^4 \lambda_\mathrm{R}^2} \,,
\label{Higgs}
\ee
where $h$ is the SM-like Higgs boson state, while the CP-odd scalar mass receives extra contributions through the \emph{portal couplings} $\lambda_{\mathrm{A} \mathrm{L}}$ and $\lambda_{\mathrm{A} \mathrm{R}}$, acquiring the form
\be
m_\mathrm{A}^2 = 2 \(v^2 \lambda_{\mathrm{AL}} + \omega_\chi^2 \lambda_{\mathrm{AR}} + \mu_\mathrm{A}^2\)\,.
\label{A-mass}
\ee

Once again, let us provide a numerical estimate, taking a purely classical field theory approach in the sense that the values of the LR theory quartic couplings are directly extracted from the trinification scalar potential at tree level. Note that both tree-level and one-loop matching, as well as the Renormalisation Group evolution (RGE) effects, are beyond the scope of this study and will be considered in a future work. 

We first consider that the $\omega_\chi$ \vev~is developed at the same scale as $\mu_\mathrm{R}$. Thus we fix it to
\be
\omega_\chi = 9~\text{TeV} \qquad
\text{while}\qquad v = 246~\text{GeV} \,.
\label{VEVs}
\ee
The quartic couplings of the LR theory, at our level of accuracy, depend solely on the $\kappa_1^{\(\chi\)}$, $\kappa_1^{\prime \(\chi\)}$ and $\kappa_{12}^{\(\chi\)}$, as well as on the scalar mixing angles of the trinification theory. Using Eq.~\eqref{quartics-num} we get
\be
\lambda_\mathrm{L} \approx 0.164 \,,
\qquad
\lambda_{\mathrm{R}} \approx 0.135 \,,
\qquad
\lambda_{\mathrm{L} \mathrm{R}} \approx 0.137 \,,
\qquad
\lambda_{\mathrm{A} \mathrm{L}} \approx \lambda_{\mathrm{A} \mathrm{R}} \approx 0.068\,.
\label{matching-num}
\ee
Taking $\mu_\mathrm{A}^2 = \mu_{(\chi)12}^2$ in this example, the scalar masses become
\be
m_h \approx 125~\mathrm{GeV} \,,
\qquad
m_H \approx 4.7~\mathrm{TeV} \,,
\qquad
m_\mathrm{A} \approx 184~\mathrm{GeV}\,,
\ee
suggesting that our model is compatible with the Higgs sector of the SM, and offers a new heavy CP-even scalar as well as a Majoron state.

\begin{figure}[htb]
\centering
\begin{center}
\includegraphics[height=6cm,width=9cm]{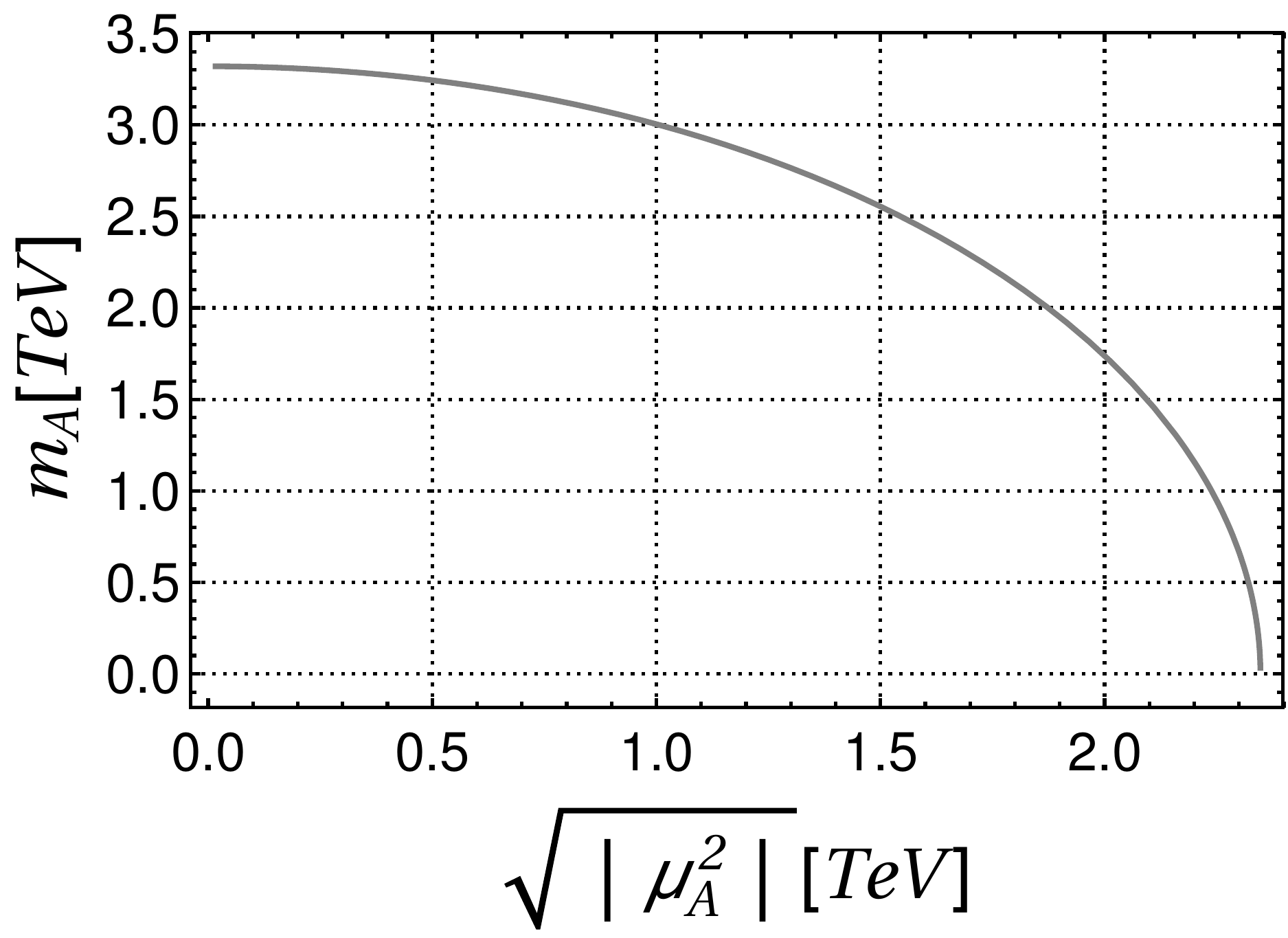}
 \end{center}
\caption{\footnotesize The Majoron mass as a function of the size of the accidental 
$\U{acc}$ soft-breaking term.}
\label{fig:mA}
\end{figure}

Note that different choices for the size of the $\mathbb{Z}_2^{(1)}$ and $\U{acc}$ soft breaking parameter $\mu_{(\chi)12}^2$ yield distinct Majoron masses. We show in Fig.~\ref{fig:mA} the allowed values of $m_\mathrm{A}$ as a function of the size of the soft-breaking parameter $\mu_\mathrm{A}$, while keeping all other parameters fixed as in the example above.

Note that the numerical example that we have outlined above is simply indicative of the key properties of the model, and a full phenomenological analysis is left for a future work. Let us also mention that, in addition to the SM-like gauge bosons, the model also contains new $W^\prime$ and $Z^\prime$ gauge bosons. Their masses can be either at the $\omega_\chi$ scale, if the corresponding gauge couplings are of order unity, or at the TeV scale if such gauge couplings are of order $0.1$. Therefore, the gauge sector of our model also offers interesting prospects for the LHC Run-III, which is scheduled to start in 2021.

While a detailed analysis of the FCNC constraints goes beyond the scope of the current work,
we can make a generic statement about non-existence of the tree-level FCNCs in our model based
upon the Glashow-Weinberg-Paschos theorem \cite{Glashow:1976nt,Paschos:1976ay}. This theorem
states that there will be no tree-level FCNC processes coming from the scalar sector if all
right-handed fermions of a given electric charge couple to only one of the L-doublets. As was 
demonstrated above the minimal low-energy LR $\SU{2}{L} \times \SU{2}{R} \times \U{B-L}$ symmetric 
scenario in the considered trinification model features the scalar sector composed of one $\SU{2}{L}$ 
doublet, one $\SU{2}{R}$ doublet and one pseudo-Goldstone state. As follows from from Eqs.~(\ref{Lq}) 
and (\ref{Ll}), the condition of the Glashow-Weinberg-Paschos theorem is automatically satisfied 
in this case. Possible FCNC contributions would emerge at loop level only rendering the model safe 
with respect to the corresponding phenomenological constraints.

\section{Gauge Coupling Unification}
\label{Sec:GCU}

In this section, we study the RGE to determine if the couplings of the $\SU{3}{C}$, $\SU{3}{L}$, $\SU{3}{R}$ and $\U{X}$ gauge groups unify 
at some high scale $M_U$. In our analysis, we take into account the intermediate scale (denoted as $M_2$) in Eq.~(\ref{SB}), so that in the energy 
ranges $\mu\leqslant M_2$, $M_2\leqslant \mu\leqslant M_3$, $M_3\leqslant \mu\leqslant M_U$ the theory is described by the gauge groups 
$\SU{3}{C}\times \SU{2}{L}\times \U{Y}$, $\SU{3}{C}\times \SU{2}{L}\times \SU{2}{R}\times \U{B-L}$ and $\SU{3}{C}\times \SU{3}{L}\times \SU{3}{R}\times \U{X}$, 
respectively. Here, $M_2$ and $M_3$ are the scales where the gauge groups $\SU{3}{C}\times \SU{2}{L}\times \SU{2}{R}\times \U{B-L}$ and 
$\SU{3}{C}\times \SU{3}{L}\times \SU{3}{R}\times \U{X}$ are spontaneously broken, respectively. Besides that, in the energy range 
$M_2\leqslant \mu\leqslant M_3$, for the RGE analysis we consider that the relevant scalar content is the one corresponding 
to the next-to-minimal light scalar sector discussed in the previous section.

The starting point of our RGE analysis is the one-loop RG equation for a given structure constant $\alpha_{i}=g_{i}^{2}/(2\pi)$ (with $g_{i}$ being 
the respective gauge coupling) which is given by: 
\begin{equation}
\frac{d\alpha _{i}(\mu )}{dt}=-\frac{b_{i}^{\rm eff}(\mu )}{2\pi } \,, \qquad t=\ln \left( \frac{\mu }{\mu _{0}}\right) \,.
\end{equation}
It provides the running of the inverse structure constants at one-loop level as follows:
\begin{equation}
\alpha _{i}^{-1}(\mu )=\alpha _{i}^{-1}(\mu _{0})-\frac{b_{i}^{\rm eff}(\mu )}{%
2\pi }\ln \left( \frac{\mu }{\mu _{0}}\right) \,,  \label{GUT4}
\end{equation}%
The effective one-loop $\beta$-function coefficients, taking into account the thresholds
from particles with masses $m_{f}$, are given by 
\begin{equation}
b_{i}^{\rm eff}(\mu )=\sum_{f}\theta (\mu -m_{f})b_{i}^{f} \,.
\end{equation} 
The contribution of each particle $b_{i}^{f}$ is calculated according to 
\begin{equation}
b_{i}=-\frac{11}{3}T_{i}(R_{G})+\frac{2}{3}T_{i}(R_{F})+\frac{1}{3}T_{i}(R_{S}) \,,
\end{equation}%
where $T(R_{I})$ are the Dynkin indices of the representations $R_{I}$ to
which the gauge bosons $I=G$, fermions $F$, and scalars $S$, respectively, belong. 
They are defined as $T(R)\delta _{mn}=\mathop{\rm Tr}\nolimits(T_{m}(R)T_{n}(R))$, 
with $T_{m}(R)$ being the generators in the representation $R$. For the lowest-dimension 
representations of $\SU{N}{}$, they are $T_F=1/2$, $T_A=N$, for fundamental and adjoint 
representations, respectively.

Considering that the electric charge in our 3331 model is defined according to (\ref{eq:def-Q}) and (\ref{eq:def-beta}),
 where the $\SU{3}{L,R}$ generators have the normalization ${\rm Tr}\left(T_{i}^{\left( L,R\right) }T_{j}^{\left( L,R\right) }\right) =
\frac{1}{2}\delta _{ij}$, we can define normalized operators $\left( B-L\right) _{N}$, $X_{N}$ and $Y_{N}$ satisfying the relations:
\begin{equation}
\left( B-L\right) =n_{B-L}\left( B-L\right) _{N},\hspace{1.5cm}X=n_{X}X_{N} \,, \hspace{1.5cm} 
Y=n_{Y}Y_{N} \,.
\end{equation}
The normalization factors $n_{B-L}$, $n_{X}$ and $n_{Y}$ fulfil the relations:
\begin{equation}
n_{B-L}^{2}=\frac{4}{3}+4n_{X}^{2},\hspace{1.5cm}n_{Y}^{2}=1+\frac{1}{4}
n_{B-L}^{2}=\frac{4}{3}+n_{X}^{2} \,.
\end{equation}
Here we have taken into account that in the low-energy LR symmetric theory 
the hypercharge operator is defined as follows:
\begin{equation}
Y=T_{2R}+\frac{1}{2}\left( B-L\right) \,,
\end{equation}
The normalized couplings are related by the following matching conditions:
\begin{equation}
\left( \frac{1}{3}+n_{X}^{2}\right) \left( \alpha _{B-L}^{N}(M_{3})\right)
^{-1}=\frac{1}{3}\left( \alpha _{3L}^{-1}(M_{3})+\alpha
_{3R}^{-1}(M_{3})\right) +n_{X}^{2}\left( \alpha _{X}^{N}(M_{3})\right) ^{-1} \,,
\end{equation}
\begin{equation}
\left( \frac{4}{3}+n_{X}^{2}\right) \left( \alpha _{Y}^{N}(M_{2})\right)
^{-1}=\alpha _{2R}^{-1}(M_{2})+\left( \frac{1}{3}+n_{X}^{2}\right) \left(
\alpha _{B-L}^{N}(M_{2})\right) ^{-1} \,.
\end{equation}
From the embedding of $\SU{2}{L,R}$ into $\SU{3}{L,R}$ we have the following additional matching condition:
\begin{equation}
\alpha _{3\left( L,R\right) }\left( M_{3}\right) =\alpha _{2\left(L,R\right) }\left( M_{3}\right) \,.
\end{equation}
Furthermore, we have the following relations: 
\begin{equation}
\alpha_{X}^{N}=n_{X}^{2}\alpha_{X},\hspace{1.5cm}\alpha _{B-L}^{N}=4\left( \frac{1}{3}+n_{X}^{2}\right) \alpha_{B-L} \,.
\end{equation}
\begin{equation}
\alpha _{Y}^{N}=n_{Y}^{2}\alpha_{Y}=\left( 1+\frac{1}{4}n_{B-L}^{2}\right)\alpha _{Y}=\left( \frac{4}{3}+n_{X}^{2}\right) \alpha_{Y} \,.
\end{equation}
In addition, we require the following condition:
\begin{equation}\label{eq:kappa}
\alpha _{2R}^{-1}(M_{2})=\kappa \alpha _{2L}^{-1}(M_{2}) \,,
\end{equation}
where $\kappa$ is an $\mathcal{O}(1)$ parameter.

Using the relations and considerations described above, we find that the inverse 
structure constants of the $\SU{3}{C}$, $\SU{3}{L}$, $\SU{3}{R}$ and $\U{X}$ gauge groups 
evaluated at the energy scale $\mu\geqslant M_3$ are given by:
\begin{equation}\label{eq.aC}
\alpha _{3C}^{-1}(\mu )=\alpha _{3C}^{-1}(M_{Z})-
\frac{b_{\SU{3}{C}}^{\left( SM\right) }}{2\pi }\ln \left( \frac{M_{2}}{M_{Z}}\right) -
\frac{b_{\SU{3}{C}}^{\left( 3221\right) }}{2\pi }\ln \left( 
\frac{M_{3}}{M_{2}}\right) -\frac{b_{\SU{3}{C}}^{\left(
3331\right) }}{2\pi }\ln \left( \frac{\mu }{M_{3}}\right) \,,
\end{equation}
\begin{equation}\label{eq.aL}
\alpha _{3L}^{-1}(\mu )=\alpha _{2L}^{-1}(M_{Z})-
\frac{b_{\SU{2}{L}}^{\left( SM\right) }}{2\pi }\ln \left( \frac{M_{2}}{M_{Z}}\right) -
\frac{b_{\SU{2}{L}}^{\left( 3221\right) }}{2\pi }\ln \left( 
\frac{M_{3}}{M_{2}}\right) -\frac{b_{\SU{3}{L}}}{2\pi }\ln
\left( \frac{\mu }{M_{3}}\right) \,,
\end{equation}
\begin{equation}\label{eq.aR}
\alpha _{3R}^{-1}(\mu )=\alpha _{2L}^{-1}(M_{Z})-
\frac{b_{\SU{2}{L}}^{\left( SM\right) }}{2\pi }\ln \left( \frac{M_{2}}{M_{Z}}\right) -
\frac{b_{\SU{2}{R}}}{2\pi }\ln \left( \frac{M_{3}}{M_{2}}
\right) -\frac{b_{\SU{3}{R}}}{2\pi }\ln \left( \frac{\mu }{M_{3}
}\right) \,,
\end{equation}
\begin{eqnarray}\label{eq.aX}
\alpha _{X}^{-1}(\mu ) &=&\alpha _{Y}^{-1}(M_{Z})-
\frac{b_{\U{Y}}}{2\pi }\ln \left( \frac{M_{2}}{M_{Z}}\right) -\kappa \alpha
_{2L}^{-1}(M_{Z})+\frac{\kappa b_{\SU{2}{L}}^{\left( SM\right) }
}{2\pi }\ln \left( \frac{M_{2}}{M_{Z}}\right)\notag  \\
&&-\frac{1}{4}\times \frac{b_{\U{B-L}}}{2\pi }\ln \left( 
\frac{M_{3}}{M_{2}}\right) -\frac{1}{3}\left[ \left( \kappa +1\right) \alpha
_{2L}^{-1}(M_{2})-\frac{b_{\SU{2}{L}}^{\left( 3221\right)
}+b_{\SU{2}{R}}}{2\pi }\ln \left( \frac{M_{3}}{M_{2}}\right) 
\right]\notag \\
&&-\frac{b_{\U{X}}}{2\pi }\ln \left( \frac{\mu }{M_{3}}
\right) \,,
\end{eqnarray}
where
\begin{eqnarray}
b_{\SU{3}{C}}^{\left( 3331\right) } &=&-5 \,,\hspace{1.5cm}
b_{\SU{3}{L}}=-\frac{16}{3}\,,\hspace{1.5cm}
b_{\SU{3}{R}}=-\frac{11}{2}\,,\hspace{1.5cm}
b_{\U{X}}=\frac{28}{9} \,, 
\notag \\
b_{\SU{3}{C}}^{\left( 3221\right) } &=&-7\,,\hspace{1.5cm}
b_{\SU{2}{L}}=-\frac{7}{3}\,,\hspace{1.5cm}
b_{\SU{2}{R}}=-\frac{19}{6}\,,\hspace{1.5cm}
b_{\U{B-L}}=\frac{16}{9}\,, 
\notag \\
b_{\SU{3}{C}}^{\left( SM\right) } &=&-7\,,\hspace{1.5cm}
b_{\SU{2}{L}}^{\left( SM\right) }=-\frac{19}{6}\,,\hspace{1.5cm}
b_{\U{Y}}=\frac{41}{10}\,.
\end{eqnarray}
Requiring that the couplings of the $\SU{3}{C}$, $\SU{3}{L}$, $\SU{3}{R}$ and $\U{X}$ gauge groups 
unify at some high scale $M_U$, implies the following condition:
\begin{equation}\label{eq:Uni}
\alpha _{3C}^{-1}(M_{U})=\alpha _{3L}^{-1}(M_{U})=\alpha
_{3R}^{-1}(M_{U})=\left( \alpha _{X}^{N}(M_{U})\right) ^{-1}\,.
\end{equation}

\begin{figure}[htb]
	\centering
	\begin{center}
		\includegraphics[width=.495\textwidth]{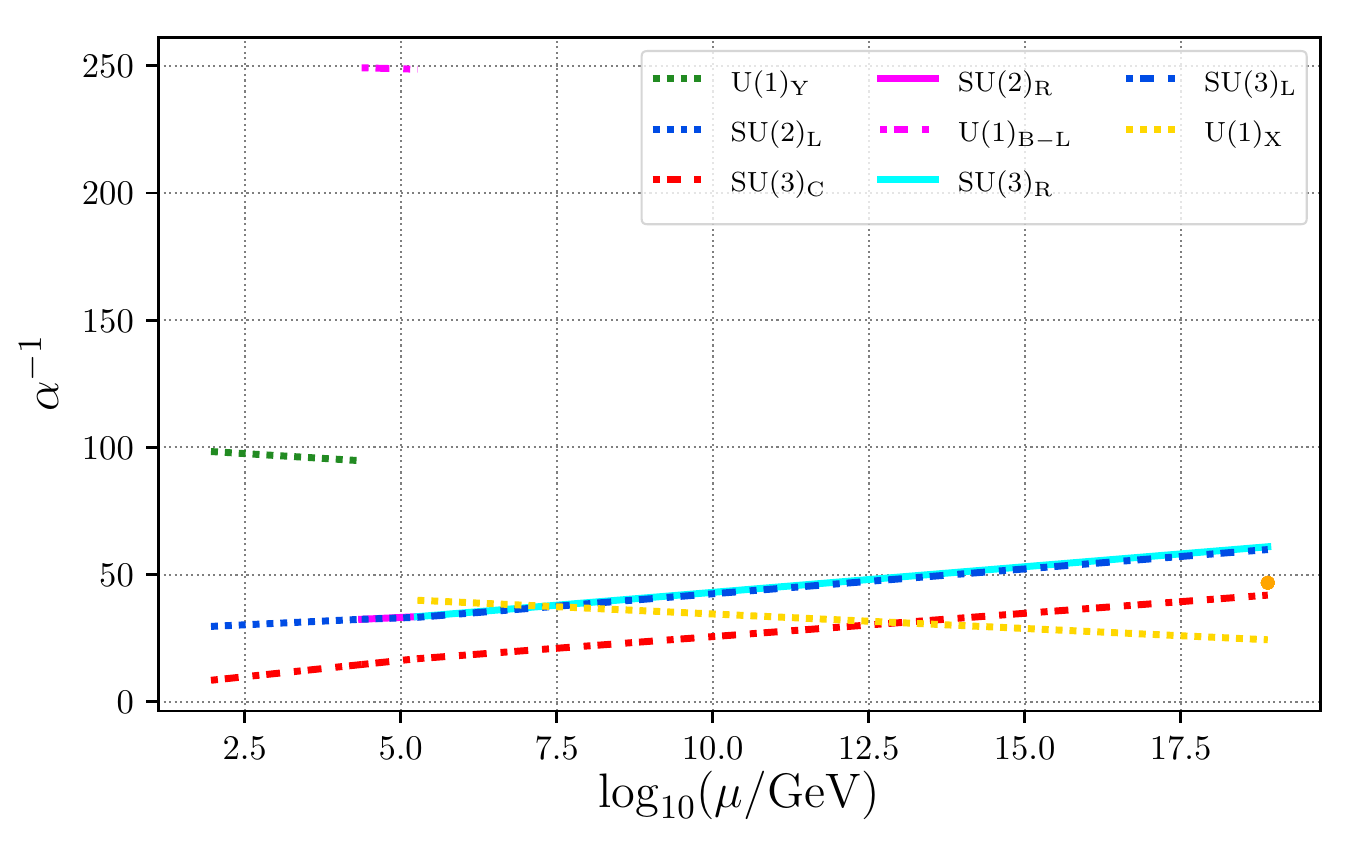}\\
		\includegraphics[width=.495\textwidth]{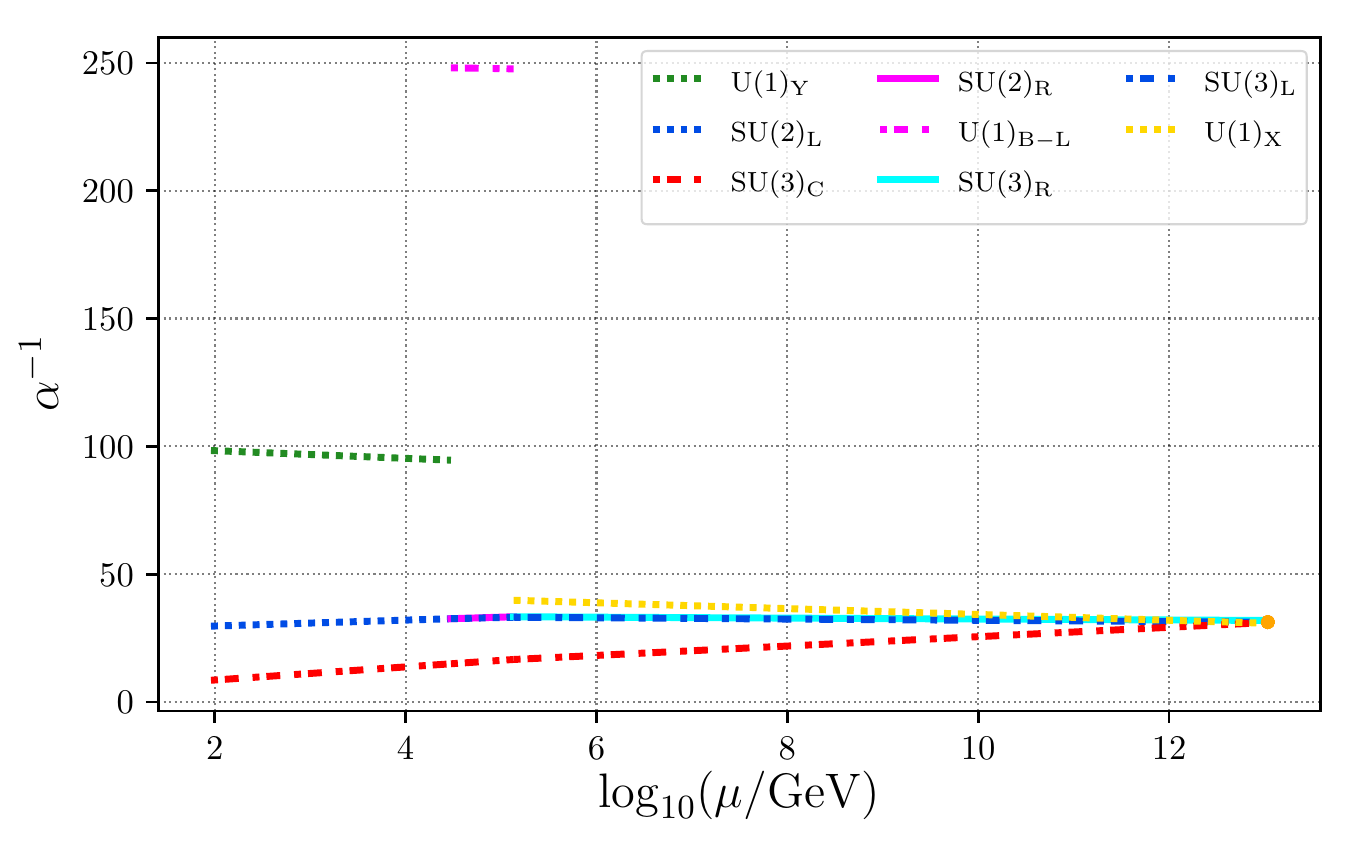}
		\includegraphics[width=.495\textwidth]{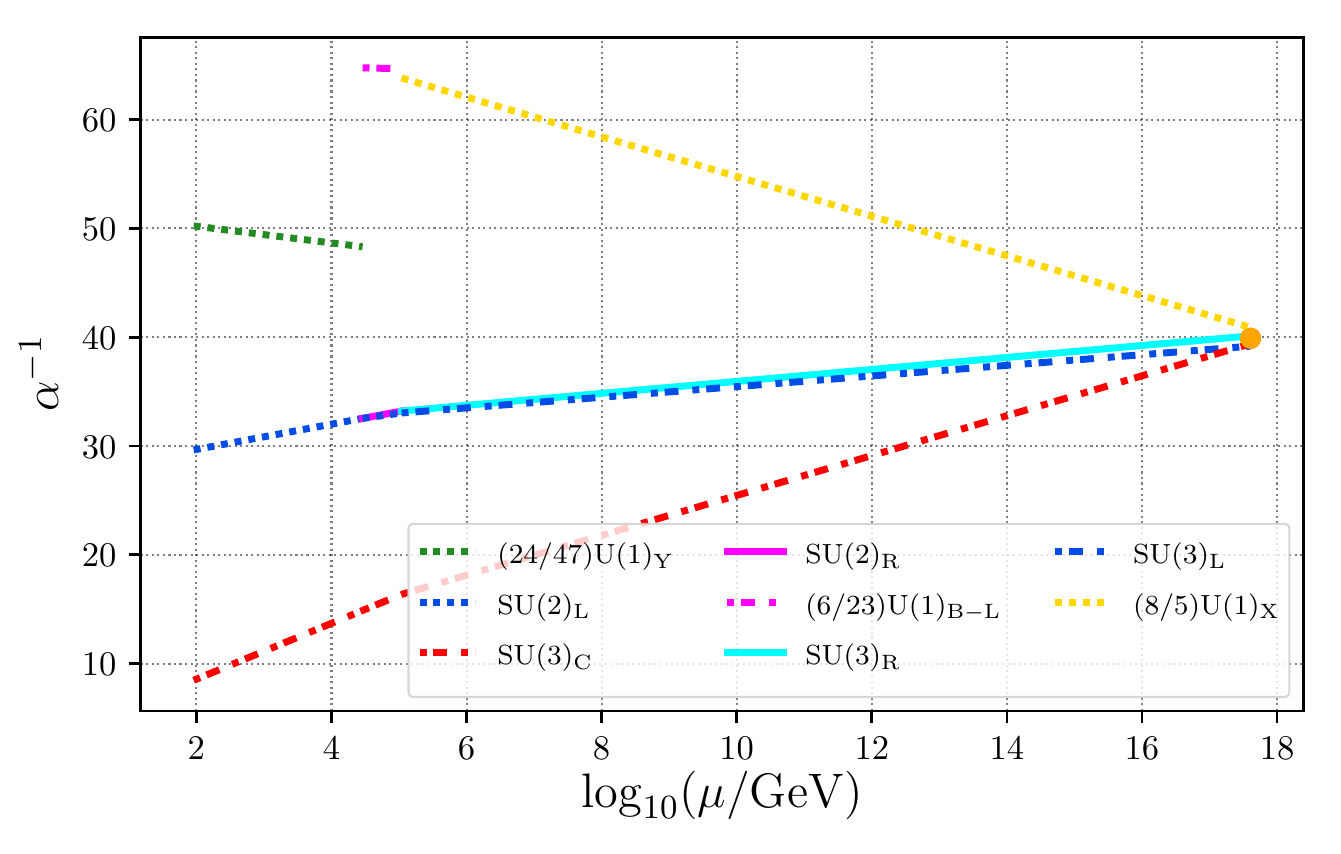}
	\end{center}
	\caption{\footnotesize Running of the gauge couplings for three benchmark scenarios. On the top panel we show the result obtained for the model discussed in this manuscript 
	while on the bottom panels we consider the inclusion of two (right) and three (left) vector-like $\SU{3}{L} \times \SU{3}{R}$ fermion bi-triplets. The orange dots represent the average 
	value of $\alpha _{3L}^{-1}(M_U )$, $\alpha _{3R}^{-1}(M_U )$,  $\alpha _{3C}^{-1}(M_U )$ and $\alpha _{X}^{-1}(M_U )$ and are merely indicative of the optimal unification point. 
	While on the top and bottom-left panels the charge normalization factors read $n_{X}^{2} = n_{B-L}^{2} = n_{Y}^{2} = 1$, on the bottom-right panel they are given by 
	$n_{X}^{2} = 5/8$, $n_{B-L}^{2} = 23/6$ and $n_{Y}^{2} = 47/24$ as indicated in the legend.}
	\label{fig:GCU}
\end{figure}

Let us then numerically examine the gauge coupling's evolution in our model and verify whether the above unification condition is achievable. 
First, we consider a scenario where beyond the scale $M_3 \sim \mathcal{O}(100~\mathrm{TeV})$ there are no additional fields on top of those 
studied in this paper and given in Tabs.~\ref{tab:scalars} and \ref{tab:fermions}. The low-scale boundary conditions are set as
\begin{equation}
	\begin{aligned}
	\alpha _{3C}^{-1}(M_Z ) &= 8.47\,, \\
	\alpha_{2L}^{-1}(M_Z) &= \alpha^{-1} \sin \theta_W^2(M_Z)\,,\\
	\alpha_{Y}^{-1}(M_Z) &= \alpha^{-1} \cos \theta_W^2(M_Z)\,,
	\end{aligned}
	\label{eq:alpha_in}
\end{equation}
with $\alpha(M_Z) = 1/128$ being the fine structure constant evaluated at the $Z$-boson mass scale and 
$\sin \theta_W^2(M_Z) = 0.232$ the weak mixing angle at the same scale \cite{Tanabashi:2018oca}.
\begin{table}[htb!]
	\begin{center}
		\begin{tabular}{ccc|c}
			\hline                     
			$\alpha^{-1}_X(M_U)$ & $M_2 [\text{TeV}]$  & $M_3 [\text{TeV}]$ & $M_U$ 	\\
			\hline
			$1 - 80$ & $10 -30$ & $100 - 200$  & $M_3 - M_\mathrm{Pl}$	\\
			\hline
		\end{tabular} 
		\caption{Scanning ranges for the numerical evaluation of the RG equations. In the last column $M_\mathrm{Pl} = 1.2 \times 10^{19}~\mathrm{GeV}$ 
		denotes the Planck scale.}
		\label{tab:scan}  
	\end{center}
\end{table} 
We have performed a scan randomly sampling the $\U{X}$ gauge coupling at the GUT scale as well as the $M_2$ and $M_3$ scales as indicated in the first three columns 
of Tab.~\ref{tab:scan}, consistently with our discussion. Using Eqs.~\eqref{eq:kappa} and \eqref{eq.aC} to \eqref{eq.aX} we obtain the high scale values of the inverse 
structure constants $\alpha _{3C}^{-1}(M_U )$, $\alpha _{3L}^{-1}(M_U )$ and $\alpha _{3R}^{-1}(M_U )$, as well as the value of $M_U$ with the restriction given 
in the last column of Tab.~\ref{tab:scan}. Note that the only allowed solution of equation (\ref{eq:kappa}) is $\kappa = 1$. The scale at which the values of the gauge couplings 
become closest (or unify) defines $M_U$. Considering for simplicity that the charge normalization factors are $n_{X}^{2} = n_{B-L}^{2} = n_{Y}^{2} = 1$, we have verified that 
while $\alpha _{3L}^{-1}(M_U )$ and $\alpha _{3R}^{-1}(M_U )$ can be unified with a precision of $2.5\%$ at $M_U \sim 10^{19}~\mathrm{GeV}$, $\alpha _{3C}^{-1}(M_U )$ is, 
at best, $37\%$ away from universality and the condition \eqref{eq:Uni} cannot be satisfactorily met. This scenario is shown on the top panel of Fig.~\ref{fig:GCU} where the orange 
dot represents the mean value of the four gauge couplings.
Note that one could argue that deviations from universality can emerge from quantum gravitational effects \cite{Chakrabortty:2008zk} or from heavy Kaluza-Klein modes when embedding 
the theory in higher dimensions \cite{Dienes:1998vg,Dienes:1998vh}. However, in the current analysis we will favour minimal solutions that can address this issue. Therefore, we will study 
whether an anomaly-free extension of the current framework with $n$-generations of neutral vector-like fermion $\SU{3}{L} \times \SU{3}{R}$ bi-triplets above $M_3$ can change the picture. 
We have then extended our scan in Tab.~\ref{tab:scan} allowing the possibility of $1$ to $6$ generations of such heavy bi-triplets. The RGE coefficients of the $\SU{3}{L}$ and $\SU{3}{R}$ 
gauge couplings are modified according to
\begin{equation}
	b_{\SU{3}{L,R}} = -\frac{16}{3} + 2 n\,.
\end{equation}
Interestingly enough we have found only two viable scenarios. The first is for $n=3$ where, up to small fluctuations, the unification scale is $M_U \approx 10^{13}~\mathrm{GeV}$ and 
where the unification condition \eqref{eq:Uni} is reproduced within $2.5\%$ for $\alpha_{U}^{-1}(M_U) \approx 31$. This class of solutions is represented in the bottom-left panel 
of Fig.~\ref{fig:GCU} with the orange dot denoting the $\left(M_U,\alpha_{U}^{-1}\right)$ pair. The second type of solutions, which we represent in the bottom-right panel 
of Fig.~\ref{fig:GCU}, feature only two generations of fermion bi-triplets, $n = 2$, and the unification of gauge couplings takes place at $ 10^{17} \lesssim M_U \lesssim 10^{18}~\mathrm{GeV}$ 
with $\alpha_U^{-1}(M_U) \approx 40$ and with the charge normalization factors $n_{X}^{2} = 5/8$, $n_{B-L}^{2} = 23/6$ and $n_{Y}^{2} = 47/24$.

With these results we have shown that a successful merging of the gauge couplings at the GUT scale needs an extension of the particle content considered in this work with either two or three generations of vector-like fermion bi-triplets. The impact of such extra states for lepton masses, in particular for the neutrino sector, is beyond the scope of the current article and is left for a future work.

\section{Dark matter}
\label{Sec:DM}

The $\mathbb{Z}_2^{(2)}$ symmetry is exact and remains unbroken by any of the \vevs~in \eqref{VEVs-tri}. Therefore, the lightest neutral 
particle which carries an odd-$\mathbb{Z}_2^{(2)}$ number can be a candidate for DM. The particles 
carrying the $\mathbb{Z}_2^{(2)}$-odd charge are the components of the scalar bi-triplets $\chi_3$ and the third quark generation, while only the neutral components
of $\chi_3$ noted as $\chi_3^{12}, \chi_3^{21}, \chi_3^{23}, \chi_3^{32}$ can potentially contain a DM candidate.

However, each component of the $\chi_3$ bi-triplet couples to a pair of quarks via Yukawa interactions, given in Eq.~(\ref{Lq}). Namely, the $\chi_3^{12}, \chi_3^{21}$ couple to a pair of light SM quarks whereas the other neutral components of $\chi_3$ bi-triplet couple to a pair of quarks including a light SM quark and an exotic 
heavy (vector-like) quark. In fact, both $\chi_3^{12}, \chi_3^{21}$ decay into a pair of light quarks and thus cannot serve as DM candidates.

In order to ensure stability of the remaining neutral components of $\chi_3$ bi-triplet, one should assume that their masses are below the mass of the exotic quarks in order to prohibit their fast tree-level decays. Then, the lightest state among $\chi_3^{23}, \chi_3^{32}$ can in principle be considered as a cold DM candidate. Indeed, its properties would then be similar to those of the scalar DM candidate discussed previously in Refs.~\cite{Cirelli:2005uq,Heeck:2015qra,Garcia-Cely:2015quu,Huong:2018ytz,LopezHonorez:2006gr,Gustafsson:2007pc,Ma:2006km,Barbieri:2006dq}. 

However, in order to get small quark-mixing angles in the 13 and 23 planes at one-loop level one should break softly the $\mathbb{Z}^{(2)}_{2}$ symmetry in the scalar sector by introducing the trilinear $f_{234} \chi_2 \chi_3 \chi_4$ interaction term. The latter interaction makes it difficult to prevent the heavy
scalar components of $\chi_3$ bi-triplet from decaying and hence
to stabilise the heavy DM candidate without a significant fine-tuning of the model parameters.

As mentioned in Sec.~(\ref{Sec:scalar-sector}), our model also predicts a CP-odd pseudo-Goldstone Majoron whose mass can 
vary greatly as shown in Fig.~(\ref{fig:mA}). Interestingly, such a state can play a role of light DM candidate under certain conditions in full analogy to the existing Majoron DM scenarios \cite{Gelmini:1983ea,Gelmini:1984pe,GonzalezGarcia:1988rw,Berezhiani:1992cd,Berezinsky:1993fm,Kachelriess:2000qc,Tomas:2001dh,Lattanzi:2007ux,Bazzocchi:2008fh}.

Indeed, starting from Yukawa interactions Lagrangian, the couplings of the Majoron to the right-handed neutrinos read
\be
-\mathcal{L}^{(l)}_Y \supset \sum_{\alpha=1}^3 \sum_{\beta=1}^3 (x_\chi)_{\alpha \beta} \bar{L}_{\alpha L} \chi_2 L_{\beta R} + {\rm h.c.} \supset i \bar{\nu}_L x_\chi \left(\frac{M_{\bar{\nu}^T \nu_L^c}}{M_S}\right)^2  \nu_R A + {\rm h.c.} \,.
\ee
In the seesaw regime, the Majoron can decay into the light 
neutrinos with partial widths proportional to $m_j^4$,
\be
\Gamma(A \rightarrow \nu \nu) \simeq \frac{m_A}{8 \pi f^4} \sum_{j=1}^3 m_j^4 \simeq \frac{1}{3 \times 10^{22}s} \left( \frac{m_A}{1 {\rm MeV}} \right) \left(\frac{10^9 {\rm GeV}}{f}\right)^4 \left( \frac{\sum_{i=1}^3 m_i^4}{10^{-6} {\rm eV}^4}\right) \,, \quad f \equiv {\rm Tr}\Big[\frac{M_S}{\sqrt{x_\chi}}\Big] \,.
\ee
It is straightforward to notice that a light Majoron can easily be long-lived enough to be a DM candidate for typical seesaw scales, 
assuming that $A \rightarrow \nu \nu$ is the main Majoron 
decay channel. 

Just as in the singlet Majoron model \cite{Chikashige:1980ui,Pilaftsis:1993af}, the Majoron couples 
also to the charged fermions, $g_{A\bar{f}f}$ through the EW one-loop diagrams, due to the mixing between the new neutral fermions, $N_a$, and the active neutrinos $\nu_a$. The coupling to quarks is induced by a one-loop $A-Z^0$ mixing with neutrinos being inside the loop, and the coupling to the charged leptons is obtained by an analogue $Z^0$ exchange diagram and additional $W$ exchange graphs, see Refs.~\cite{Garcia-Cely:2017oco,Chikashige:1980ui}. Due to a small coupling of the Majoron to neutrinos, which is suppressed by $(M_{\bar{\nu}^T \nu_L^c}/M_S)^2$, these diagrams give a rather small contribution to the coupling strength $g_{All}$ \cite{Chikashige:1980ui}. However, the Majoron couples to the exotic quarks and neutral fermions $N_a$ that interact with new heavy gauge bosons, $Z^\prime_{L,R}, Z_R, X^{0,0*}_{L,R}, Y^\pm_{L, R}$. The additional diagrams of the Majoron coupling to the fermions are predicted such as the $A-Z_{LR}^\prime, A-Z_R$ mixing graphs with $N_a$ in the loop and the graphs that are mediated by the exotic quarks and the $Y^\pm$ gauge boson are suppressed by a factor $m_{W}^4/M_{Y_{L,R}}^4$. If the new physics scale is of the order of $100$ TeV, then the contribution to the effective coupling of the Majoron to the charged leptons is not small enough to ensure that the Majoron's lifetime is larger than the age of the Universe. Therefore, to prevent the Majoron decays into charged fermions, we need to impose an upper limit on the Majoron mass $m_A < 2 m_e \sim 1$ MeV yielding a tantalising possibility for warm DM in our model.

The Majoron would only be considered as a successful DM candidate if its relic density is consistent with cosmological observations \cite{Komatsu:2008hk}. In this sense, the coupling of the Majoron to the SM Higgs plays an important role in order to determine the DM relic density \cite{Silveira:1985rk,Garcia-Cely:2017oco,Frigerio:2011in}. In the considered model, the Majoron has a quartic coupling with the SM Higgs boson,
\be
V_{\rm LR} \supset \lambda_{\rm LR}A^2 L^\dag L \supset \lambda_{\rm LR}vA^2 h + \frac{\lambda_{\rm LR}}{2}A^2 h^2 \,.
\ee
Thus, the light Majoron can be produced by the SM Higgs decay, 
$h \rightarrow AA$. The corresponding decay rate is given by \cite{McDonald:1993ex}
\be
\Gamma(h \rightarrow AA)=\frac{1}{16 \pi}\lambda_{\rm LR}^2 \frac{v^2}{m_h}\sqrt{1-4\frac{m_A^2}{m_h^2}} \,.
\ee
There are two production mechanisms for DM known as the freeze-out and freeze-in mechanisms. The Majoron cannot be produced by the freeze-out mechanism due to strong constraints from the direct detection measurements and the LHC bounds on the invisible decay of the SM Higgs \cite{Garcia-Cely:2017oco,Frigerio:2011in,Hall:2009bx}, while the freeze-in mechanism can efficiently produce the correct DM density. For such a scenario, the Majoron relic density is determined by
\be
\Omega_A h^2 \simeq 2 \frac{1.09 \times 10^{27}}{g_*^s \sqrt{g_*^\rho}} \frac{m_A\Gamma(h \rightarrow AA)}{m_h^2} \,,
\label{DMrelic}
\ee
where $g^s_*$ and $g_*^\rho$ are the numbers of degrees of freedom contributing to the entropy and energy densities when the Majoron decouples. To obtain the corrected relic density given by Ref.~\cite{Komatsu:2008hk}, we can derive the constraint from Eq.~(\ref{DMrelic}) as follows 
\be
\lambda_{\rm LR}\simeq  2 \times 10^{-10}\sqrt{\frac{1 {\rm MeV}}{m_A}} > 2 \pi \times 10^{-9} \,.
\ee

Note that with a quartic interaction of the Majoron with a SM-like Higgs doublet the possibility for collider searches of DM in the invisible Higgs decay channel is opened. In this case, the DM signature can emerge as missing energy in the production processes at the LHC. Another possibility is via indirect DM detection channels through the relic Majoron scattering off nucleons via $t$-channel exchange of the SM Higgs boson. For more detail, see Refs.~\cite{Silveira:1985rk,McDonald:1993ex}.

On the other hand, the Majoron couples to two gauge bosons via two-loop diagrams. A detailed analysis of these two-loop contributions has not been performed in this work. However, based upon the results given in Ref.~\cite{Garcia-Cely:2017oco} and the new contributions to the effective one-loop couplings specific to the considered model, we estimate the coupling of the Majoron to photons to be very small. This implies that the estimated decay rate $A \rightarrow \gamma \gamma$ is more suppressed than the corresponding decay into a neutrino pair, $A \rightarrow \nu \nu$. We conclude that effective Majoron-photon coupling is consistent with astrophysical limits \cite{Yuksel:2007dr} in the considered case of light Majoron, $m_A < 1$ MeV.

\section{Leptogenesis}
\label{Sec:Leptogenesis}

In our model, both left-handed and right-handed neutrinos carry one unit of $B-L$ charge and acquire Majorana masses via a radiative correction after the spontaneous $\U{B-L}$ breaking. 
It constitutes a source for lepton asymmetries, which must be produced entirely during or after the $B-L$ symmetry breaking stage. Therefore, the lepton asymmetry can realize due to 
CP-violating decays of the sterile neutrinos. The relevant Yukawa interactions are given by
\be
-\mathcal{L}_{Y}\supset \sum_{\alpha} \sum_{\beta} \left(x_\chi \right)_{\alpha \beta} \overline{L}_{\alpha L} \chi_2 L_{\beta R} \supset 
(x_\chi)_{\alpha \beta}\bar{e}_{\alpha L}\nu_{\beta R} (\chi_2)_{21}^- + (x_\chi)_{\alpha \beta}\bar{v}_{\alpha L}N_{\beta R} (\chi_2)_{13}+{\rm h.c.} \,,
\label{LFVi}
\ee
The neutral scalar state, $(\chi_2)_{13}$, carries one unit of lepton number, while the neutral leptons, $N_{a R,L}$, do not have a lepton number and 
the left- and right-handed neutrinos, $\nu_{\beta L,R}$, acquire Majorana mass. Thus, the interactions given in Eq.~(\ref{LFVi}) are indeed 
lepton number violating. 

The lepton asymmetry can then be created in two possible ways. One way is via decays of the right-handed Majorana neutrinos as:
\be
\nu_{aR} \rightarrow e_{bL} (\chi_2^+)_{12} \,, 
\ee 
and another way is via the decay of sterile neutrino $N_{aR}$:
\be
N_{aR} \rightarrow \nu_{bL} (\chi_2)_{13} \,.
\ee
The right-handed neutrinos couple to the heavy right-handed charged gauge boson $W_R^\pm$ giving rise to stringent constraints on the $W_R$ mass from 
the out of equilibrium dynamics of the right-handed neutrino and washing out of the lepton asymmetry \cite{LRL, LRL1, LRL2}. Particularly, the scattering process 
$e^{\pm}_{R}W_{R}^{\pm} \rightarrow \nu_{aR}(H^{\pm \pm}) \rightarrow e^{\mp}_R W_R^{\mp}$ will be in equilibrium until some temperature close to the EW 
phase transition and continue to washout the asymmetry if $W_R^\pm$ with mass in the $\text{TeV}$ range \cite{LRL2}. Thus, the successful high-scale leptogeneis, 
which is associated with the decay of right-handed neutrinos, requires the mass of $W_R^\pm$ to be large \cite{LRL,LRL2}. In other words, if the new physics scale 
of the model is in the range of a few hundred TeV or somewhat larger, the lepton flavor violating decays of right-handed neutrinos do not contribute to the amount 
of baryon asymmetry in the universe.

Noting that the remaining sterile neutrino, $N_{aR}$, couples to new gauge bosons as follows
\be
\mathcal{L}^{c.c}\supset -i\frac{g_R}{\sqrt{2}}\left(\overline{N_{aR}^c}Y^{+}_{\mu_R}\partial^{\mu}e_{aR}+\overline{N_{aR}^c}X_{\mu_R}^0\partial^{\mu}\nu_{aR} \right)+h.c..
\label{gauge1}
\ee
Depending on the mass hierarchy between $N_{aR}$ and $Y^\pm_\mu$, the first term in Eq.~(\ref{gauge1}) provides the scattering processes 
$e^{\pm}_Re^\pm_R \rightarrow N_{aR} \rightarrow Y^{\pm}_R Y^\pm_R$ or $e^{\pm}_{R}Y_{R}^{\pm} \rightarrow N_{aR} \rightarrow e^{\mp}_R Y_R^{\mp}$. 
If $m_{N_R}> m_{Y^\pm}$, the first scattering is allowed and it goes out-of-equilibrium if $m_{N_R} > 10^{16}\, \text{GeV}$. In the case $m_{N_R}< m_{Y^\pm}$, 
leptogenesis occurs either at $T> m_{Y_R}$ or at $T=m_{Y_R}$, and the condition for second scattering process going out-of-equilibrium reads: 
\begin{eqnarray}
m_{Y^\pm}\geq 3 \times 10^6 \left( \frac{m_{N_R}}{10^2\, \text{GeV}} \right)^{\frac{2}{3}}\, \text{GeV} \,.
\end{eqnarray}
The second term in Eq.~(\ref{gauge1}) is responsible for $X^0_R+X^{0*}_R \rightarrow \nu_R +\nu_R$ scattering, but it is less important since the right-handed neutrino 
is a heavy particle. Noting that the new gauge bosons $X_{R}^{0}, Y_{R}^\pm$ carry one unit of lepton number, the interactions given in Eq.~(\ref{gauge1}) are lepton number 
conserving. The above gauge boson scattering processes are not efficient in washing out the lepton asymmetry. Therefore, if we assume that $m_{N_R}< m_{Y^\pm}$, leptogenesis 
occurs at temperature satisfying either $T=m_{Y_R}$ or $T> m_{Y_R}$, when $m_{N_{aR}} \simeq \mathcal{O}(1)\, \text{TeV}$, such that one obtains a lower bound 
$m_{Y_R}> 10^4 \, \text{TeV}.$ It means that in order to achieve successful leptogenesis, the $\SU{3}{R}$ breaking scale must be greater than $10^4\, \text{TeV}.$ 

Let us consider now an amount of the baryon asymmetry produced from decays of sterile neutrino $N_{aR}$ with an assumption $m_{N_R}< m_{Y^\pm}$. We assume 
a normal mass hierarchy for the heavy right-handed sterile neutrinos, thus implying that the final lepton asymmetry is given only by the CP-violating decay of the lightest one, 
$(N_{1R})$. The CP-asymmetry $\epsilon_1$ comes from a superposition of tree-level contribution, self-energy correction, and the one-loop radiative corrections 
via diagrams involving the heavier Majorana neutrinos $N_{2R}, N_{3R}$. Thus, it can be written as follows
\be
	\epsilon_1 =\frac{1}{16 \pi (x_\chi^{\dag} x_\chi)_{11}}\sum_{j \neq 1} \Im\[(x_{\chi}^{\dag} x_\chi)^2_{1j}\] g(\xi_{j1}) \,,
\ee 
where $\xi_{j1}=m_{N_{jR}}^2/m_{N_{1R}}^2$, and
\be
	g(\xi)=\sqrt{\xi}\left[ \frac{2}{1-\xi}+1-(1+\xi)\ln \frac{1+ \xi}{\xi}\right] \,.
\ee
We would like to note that 
\be
	\sum_{j \neq k}\Im\[(x_{\chi}^{\dag} x_\chi)^2_{1j}\]\sqrt{\xi_{j1}^\prime}= \frac{ \kappa}{ m_{\nu_{1R}}} \sum_{\alpha, \beta}
	\Im \[(x_{\chi}^*)_{\alpha 1}(x_{\chi}^*)_{\beta 1}M_{\bar{\nu}_{\alpha L}\nu_{\beta R}}\]
\ee
with an assumption that
\be 
\kappa=16 \pi^2 \frac{m_{{\nu_{2R}}}}{m_{N_{2}} f(m_{N_2},m^2_{{\rm Re}(\chi_2)_{13}},m^2_{{\rm Im}(\chi_2)_{13}})}=
16 \pi^2 \frac{m_{\nu_{{3R}}}}{m_{N_{3}} f(m_{N_3},m^2_{{\rm Re}(\chi_2)_{13}},m^2_{{\rm Im}(\chi_2)_{13}})} \,, \quad 
\xi_{j1}^\prime=\frac{m_{\nu_{jR}}^2}{m_{\nu_{1R}}^2} \,.
\ee

Let us now consider the Dirac term of the neutrino mass matrix
\be
	M_{\bar{\nu}_L \nu_R}= x_\chi M_N^D x^T_{\chi} \,, \hs 
	M_N^D = {\rm Diag}(h_{N_ 1}^D,h_{N_2}^D,h_{N_3}^D )\frac{v_\chi^{(2)}}{\sqrt{2}} \,, \qquad
	h_{N_\gamma}^D = \frac{1}{16 \pi^2}(x_\chi)_\gamma f(m_{N_\gamma},m^2_{{\rm Re}(\chi_2)_{13}},m^2_{{\rm Im}(\chi_2)_{13}}) \,.
\ee
We also assume that all complex scalars acquire complex \vevs, namely, $v_\chi^{(2)}= v_\chi e^{i \theta}$. Thus, we find the diagonalizing matrices 
$U_L= O_L U^L_{\rm phase}, U_R= O_R U_{\rm phase}^R$ satisfying
\be 
	U_L^\dag M_{\bar{\nu}_L \nu_R} U_R \equiv 
	D_{m_{\bar{\nu_L} \nu_R}} \equiv 
	{\rm Diag} (m_{\nu_1}^D,m_{\nu_2}^D,m_{\nu_3}^D ) \,. 
\ee
If we choose $U^L_{\rm phase} = U^R_{\rm phase} =e^{-\frac{i \theta}{2}}$ and other couplings to be real, the matrix $ D_{m_{\bar{\nu_L} \nu_R}}$ 
can be taken real and is written as
\be
	D_{m_{\bar{\nu_L} \nu_R}}= O_L^\dag \left(x_\chi {\rm Diag}
	(h_{N_ 1}^D,h_{N_2}^D,h_{N_3}^D ) \frac{v_\chi}{\sqrt{2}} x_\chi^T\right) O_R \,.
\ee
On the other hand, we assume that $O_L O_R^\dag = {\rm Diag}(1,1,1)$, which implies that the CP-asymmetry $\epsilon_1$ can be rewritten as 
\be
	\epsilon_1 \simeq\frac{\kappa}{16 \pi M_{R_1}}\frac{\sum_{i}[(x_\chi^T O_L)_{1i}m_{\nu_i}^D (O^\dag_R x_\chi)_{i1}]}
	{\sum_{i} (x_\chi^T O_L)_{1i}(O_R^\dag x_\chi)_{i1}} \sum_{j \neq 1} \frac{g(\xi_{j1})}{\sqrt{\xi_{j1}}} \Im e^{i \theta} \,.
\ee
Therefore, the upper bound on the CP-asymmetry is given by
\be
	\epsilon_1^{\rm max} \simeq\frac{\kappa}{16 \pi m_{\nu_{1R}}}\sum_i m^D_{\nu_i}\sum_{j \neq 1} \frac{g(\xi_{j1})}{\sqrt{\xi_{j1}^\prime}} \,.
\ee
	
The lepton asymmetry is related to the observed baryon asymmetry of the universe, given in terms of the baryon number $n_b$ to entropy $s$ ratio as follows
\be
	\frac{n_b}{s}=-1.38 \times 10^{-3} \epsilon \eta \,.
\ee
Here, the efficiency factor $\eta$ measures the number density of the right-handed neutrinos with respect to the equilibrium value, the out-of-equilibrium 
condition at the decay, as well as the thermal corrections to the asymmetry. This factor depends on the effective mass,
\be
	\tilde{\rm m}=\frac{D_{m_{\overline{\nu}_L\nu_R}}D_{m_{\overline{\nu}_L\nu_R}}^T}{m_{\nu_R}}
\ee
For \~{m}$_i$ $\simeq (10^{-4}- 10^{-3})$ eV, $\eta$ can be as large as $O(10^1-10^2)$ \cite{Giudice:2003jh}. 

\begin{figure}[h]
		\begin{tabular}{cc}
			\resizebox{16cm}{8cm}{\vspace{-2cm}%
				\includegraphics{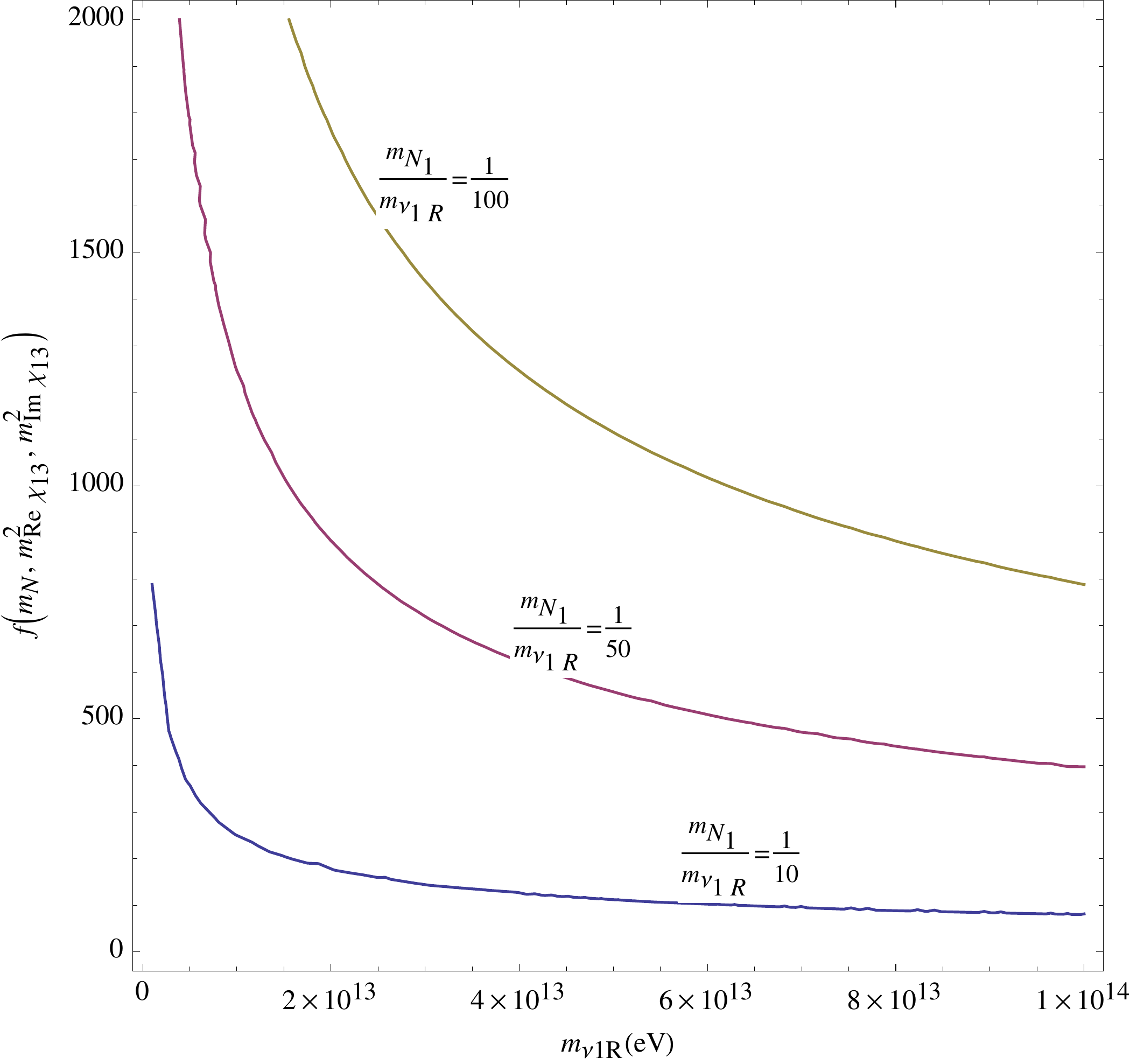}\vspace{0cm}
				\includegraphics{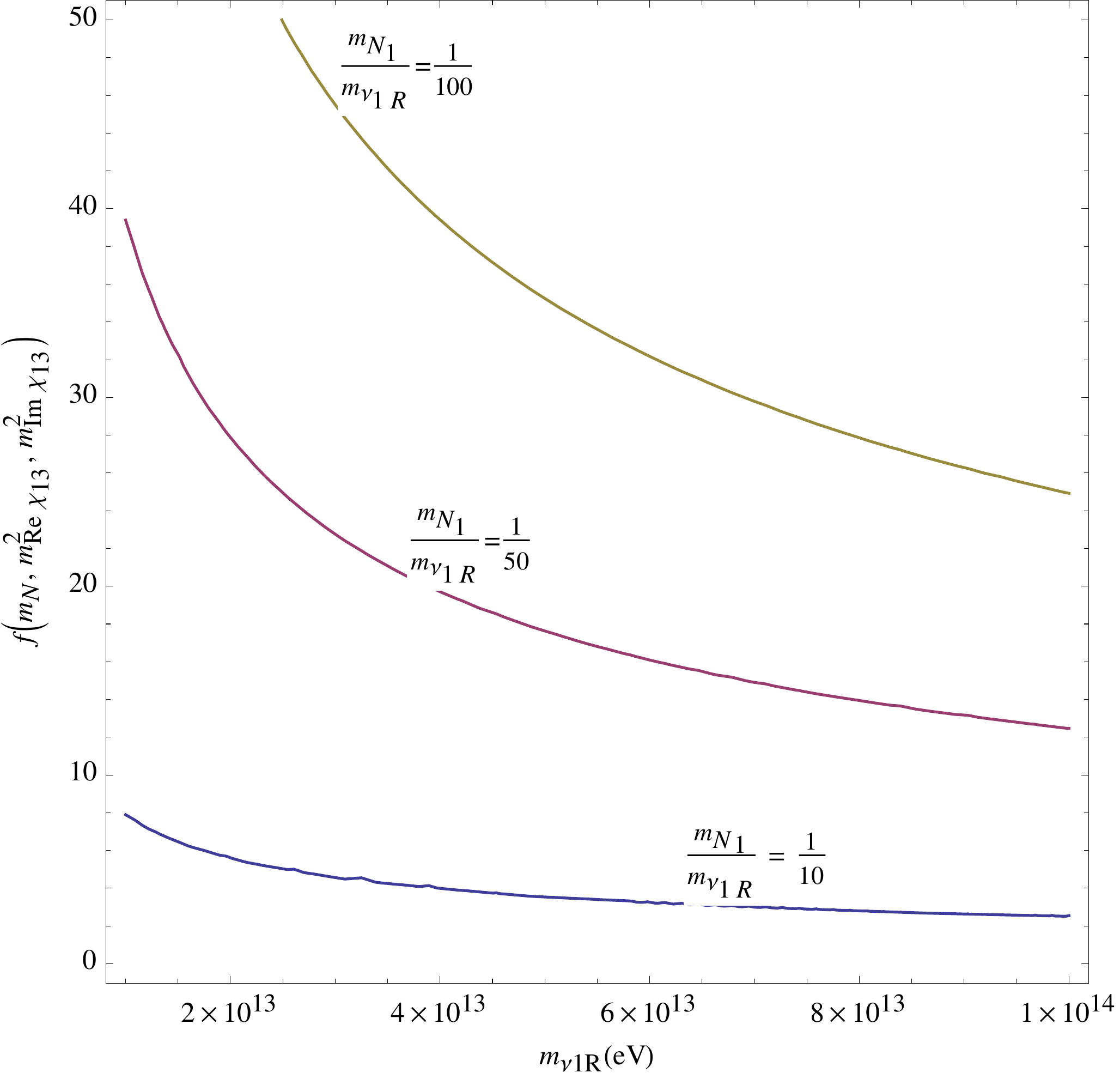}}\vspace{0cm}
		\end{tabular}
		\caption{The value of function $f(m_{N_1}, m^2_{{\rm Re}(\chi_{2})_{13}},m^2_{{\rm Im}(\chi_{2})_{13}})$ versus the lightest right-handed neutrino mass, $m_{\nu_{1R}}$, 
		for different values of the ratio $\frac{m_{N_1}}{m_{\nu_{1R}}}$, which yields the sufficient baryon number asymmetry $n_b/s = 0.87  \times 10^{-10}$. The left-handed 
		plot is obtained by taking $\tilde{\rm m}=10^{-3}, \eta=10^2$, while $\tilde{\rm m}=10^{-4}, \eta=10$ are adopted in the right-handed plot.  
		We fix $\xi^\prime_{j1}=\xi_{j1} =10$ for both cases.}
		\label{Lepto}
\end{figure}

We contour plot the correct baryon number asymmetry, $\frac{n_b}{s}=(0.87 \pm 0.04) \times 10^{-10}$, in Fig.~\ref{Lepto} in the plane of the lightest right-handed 
neutrino mass, $m_{\nu_{1R}}$, and the values of $f(m_{N_i}, m^2_{{\rm Re}(\chi_2)_{31}}, m^2_{{\rm Im}(\chi_2)_{31}})$. The allowed value of function 
$f(m_{N_i}, m^2_{{\rm Re}(\chi_2)_{31}}, m^2_{{\rm Im}(\chi_2)_{31}})$ strongly depends on the efficiency factor $\eta$, \~{m}$_i$, as well as on the ratio 
$m_{N_1}/m_{\nu_{R1}}.$ Both plots given in Fig.~\ref{Lepto} show that the allowed value of the function $f(m_{N_i}, m^2_{{\rm Re}(\chi_2)_{31}},m^2_{{\rm Im}(\chi_2)_{31}})$ 
decreases sharply as the efficiency factor $\eta$ decreases and a ratio $m_{N_1}/m_{\nu_i}$ increases. In the region $1\, \text{TeV}< m_{\nu_{1R}}< 100\, \text{TeV}$, the allowed 
value of function $f(m_{N_i}, m^2_{{\rm Re}(\chi_2)_{31}}, m^2_{{\rm Im}(\chi_2)_{31}})$ varies from a few units up to a few dozen units if $\eta=10, \tilde{\rm m}=10^{-4} $. 
This result changes hundreds of times if $\tilde{\rm m}=10^{-3}, \eta=10^2$. Noting that $\frac{m_{N_1}}{m_{\nu_{1R}}}<\frac{1}{10}$, we conclude that the sterile neutrino with mass 
in the range $\sim O(1)\, \text{TeV} - O(10)\, \text{TeV}$ can explain the baryon asymmetry via its decay.

\section{Conclusions}
\label{Sec:Conclusions}

We have built a renormalizable trinification gauge theory with an additional flavor symmetry $\U{X}\times \mathbb{Z}_{2}^{(1)} \times \mathbb{Z}_{2}^{(2)}$ at a 100 TeV energy scale, i.e.~at a much lower scale than the conventional Grand-Unified field theories imply. The low-energy spectra of this theory are shown to be consistent with the SM charged fermion mass hierarchy and the tiny values for the light active neutrino masses. Besides, the model predicts a light Majoron Dark Matter candidate in the mass range below a MeV scale and provides essential means for efficient leptogenesis. 

As the main appealing feature of the considered model, the top quark, as well as the exotic heavy fermions, obtain tree-level masses, whereas the SM charged fermions lighter than the top quark get one-loop level masses. The light active neutrino masses are generated from a combination of radiative and type-I seesaw mechanisms, with the Dirac neutrino mass matrix generated at one-loop level. The model yields one naturally light SM-like Higgs boson strongly decoupled from the other heavy scalars as well as the absence of tree-level FCNC processes mediated by the light Higgs state rendering the model safe against existing flavor physics bounds. 

The suggested flavoured trinification model can be potentially probed at the Future Circular proton-proton Collider through a discovery of ${\cal O}(10)$ TeV scale vector-like fermions, scalars and gauge bosons of trinification, while some of the next-to-lightest states
in a TeV range can also be probed by future High-Luminosity/High-Energy LHC upgrades.

\section*{Acknowledgements}
A.E.C.H, S.K. and I.S. are supported by CONICYT-Chile FONDECYT 1170803, CONICYT-Chile FONDECYT 1190845, CONICYT-Chile 
FONDECYT 1180232, CONICYT-Chile FONDECYT 3150472 and ANID PIA/APOYO AFB180002. R.P. is partially supported by the Swedish
Research Council, contract numbers 621-2013-4287 and 2016-05996, by CONICYT grant MEC80170112, as well as by the European 
Research Council (ERC) under the European Union's Horizon 2020 research and innovation programme (grant
agreement No 668679). This work was supported in part by the Ministry of Education, Youth and Sports of the Czech 
Republic, project LTT17018. A.P.M. is supported by the Center for Research and Development in Mathematics and Applications (CIDMA) through the Portuguese Foundation for Science and Technology (FCT-Fundação para a Cinência e a Tecnologia), references UIDB/04106/2020 and UIDP/04106/2020, and by national funds (OE), through FCT, I.P., in the scope of the framework contract foreseen 
in the numbers 4, 5 and 6 of the article 23, of the Decree-Law 57/2016, of August 29, changed by Law 57/2017, 
of July 19.~A.P.M.~is also supported by the projects POCI-01-0145-FEDER-022217, PTDC/FIS-PAR/31000/2017, CERN/FIS-PAR/0027/2019 and CERN/FIS-PAR/0002/2019. D.T.H acknowledges the financial support of Vietnam National Foundation for Science and Technology Development (NAFOSTED) under grant number 103.01-2019.312.

\bibliographystyle{utphys}

\bibliography{biblio}

\end{document}